%% file: lmcs08.tex
\documentclass{LMCS}

\usepackage{amsmath}
\usepackage{amssymb}
\usepackage{stmaryrd,hyperref}

\newcommand{\shortversion}[1]{}
\newcommand{\longversion}[1]{#1}

\newcommand{\para}[1]{\subsection{#1}}

\newcount\mscount
\input{./prooftree}

\usepackage{ifthen}

\newenvironment{cases*}{{\it Cases.}\hspace{2ex}}{}

\newcommand{\bla}{\ensuremath{\mbox{$$}}}
\newcommand{\der}{\,\vdash}

\newcommand{\derD}{\der}

\newcommand{\of}{\!:\!}

\newcommand{\red}{\longrightarrow}

\newcommand{\FV}{\ensuremath{\mathsf{FV}}}

\newcommand{\NN}{\mathbb{N}}

\newcommand{\defiff}{\mathrel{:\Longleftrightarrow}}

\def\lv{\mathopen{[\kern-0.2em[}}    
\def\rv{\mathclose{]\kern-0.2em]}}   
\newcommand{\den}[1]{\lv #1 \rv}
\newcommand{\Den}[2]{\den{#1}_{#2}}

\newcommand{\dens}[1]{\mathopen{[\kern-0.3ex[}#1\mathclose{]\kern-0.3ex]}}

\newcommand{\semop}[1]{\mathbin{\setlength\fboxsep{0.5\p@}\fbox{$#1$}}}

\newcommand{\bto}{\mathrel{\ensuremath{\fboxsep1pt\fbox{\raisebox{0pt}[0.5em]{$\to$}}}}}
\newcommand{\btimes}{\mathrel{\ensuremath{\fboxsep1pt\fbox{$\times$}}}}

\newcommand{\bplus}{\mathrel{\ensuremath{\fboxsep1pt\fbox{$+$}}}}

\newcommand{\bone}{\ensuremath{\fboxsep1pt\fbox{1}}}

\newcommand{\gim}{Gim\'enez}
\newcommand{\abbrev}[1]{#1}

\newcommand{\eg}{\abbrev{e.\,g.}}

\newcommand{\ie}{\abbrev{i.\,e.}}

\newcommand{\wrt}{\abbrev{w.\,r.\,t.}}

\newcommand{\etal}{\abbrev{et~al.}}

\newcommand{\Wlog}{\abbrev{W.\,l.\,o.\,g.}}

\newcommand{\loccit}{\abbrev{loc.\,cit.}}

\newcommand{\rulename}[1]{\ensuremath{\mbox{\sc#1}}}
\newcommand{\ru}{\dfrac}

\newcommand{\nru}[3]{#1\ \ru{#2}{#3}}
\newcommand{\nrux}[4]{#1\ \ru{#2}{#3}\ #4}

\newcommand{\tord}{\ensuremath{\mathsf{ord}}}

\newcommand{\s}{\mathsf{s}}

\newcommand{\forallk}[1]{\forall_{#1}}

\newcommand{\forallkind}[2][]{\ifthenelse{\equal{#1}{}}{%
  \forall #2\, }{%
  \forall #2\of#1}}
\newcommand{\forallkinddot}[2][]{\ifthenelse{\equal{#1}{}}{%
  \forall #2\!.\ }{%
  \forall #2\of#1.\ }}
\newcommand{\existskind}[2][]{\ifthenelse{\equal{#1}{}}{%
  \exists #2\, }{%
  \exists #2^{#1}}}
\newcommand{\existskinddot}[2][]{\ifthenelse{\equal{#1}{}}{%
  \exists #2\!.\ }{%
  \exists #2^{#1}\!.\ }}
\newcommand{\ofkind}[1]{\of#1}
\newcommand{\ofkinddot}[1]{\of#1.\,}

\newcommand{\nabku}[1][]{\overline\nabla\ifthenelse{\equal{#1}{}}{}{_{\!#1}}}

\newcommand{\Fpure}{\ensuremath{\mathsf{F}}}
\newcommand{\Fomega}{\ensuremath{\mathsf{F}^\omega}}
\newcommand{\Fpurehat}{\ensuremath{\mathsf{F}\kern0.6ex\widehat{}\kern0.6ex}}
\newcommand{\Fhat}{\ensuremath{\mathsf{F}_{\!\omega}\kern-0.65ex\widehat{}\kern0.6ex}}

\newcommand{\lambdahat}{\ensuremath{\lambda\,\widehat{}\;}}
\newcommand{\cxt}{\ \mathsf{cxt}}
\newcommand{\ttype}{*}
\newcommand{\ttuple}[1]{\langle #1 \rangle}

\newcommand{\tinl}{\mathsf{inl}}
\newcommand{\tinr}{\mathsf{inr}}
\newcommand{\tcaseraw}{\mathsf{case}}
\newcommand{\tcase}{\mathsf{case}}

\newcommand{\pco}{{\mathord{+}}}
\newcommand{\pcontra}{{\mathord{-}}}
\newcommand{\pin}{{\mathord{\circ}}}
\newcommand{\ptop}{\top}

\newcommand{\pinv}[1]{#1^{-1}}
\newcommand{\pupper}{\mathord{\oplus}}
\newcommand{\plower}{\mathord{\ominus}}
\newcommand{\derl}{\derD^{\vari\plower}}
\newcommand{\deru}{\derD^{\vari\pupper}}

\newcommand{\derq}{\derD^{\vari q}}

\def\tox#1{\buildrel#1\over\to}
\def\ltox#1{\buildrel\raise1pt\hbox{$\scriptstyle#1$}\over\longrightarrow}

\def\tocolow{\buildrel\raise-5pt\hbox{$\scriptscriptstyle+$}\over\rightarrow}

\newcommand{\toplus}[1]{#1\tox\pco}
\newcommand{\tominus}[1]{#1\tox\pcontra}
\newcommand{\tozero}[1]{#1\tox\pin}
\newcommand{\toco}{\tox\pco}
\newcommand{\tocontra}{\tox\pcontra}
\newcommand{\toin}{\tox\pin}

\newcommand{\pkind}{\kappa_\ttype}
\newcommand{\kfix}{{\toplus{(\toplus{\pkind}{\pkind})}{\pkind}}}
\newcommand{\muka}{\mu_{\pkind}}
\newcommand{\nuka}{\nu_{\pkind}}

\newcommand{\muinfkone}{\mu^\infty\kern-1.5ex\raisebox{-0.7ex}{\mbox{$\scriptstyle\ttype\tocolow\ttype$}}}
\newcommand{\muikone}{\mu^\vari\kern-0.6ex\raisebox{-0.5ex}{\mbox{$\scriptstyle\ttype\toco\ttype$}}}
\newcommand{\nablaikonep}{\nabla^\vari\kern-1.3ex\raisebox{-1ex}{\mbox{$\scriptstyle\ttype\tox{p}\ttype$}}}
\newcommand{\muiktwo}{\mu^\vari\kern-0.6ex\raisebox{-0.5ex}{\mbox{$\scriptstyle(\ttype\toco\ttype)\toco(\ttype\toco\ttype)$}}}

\newcommand{\nuikone}{\nu^\vari\kern-0.6ex\raisebox{-0.5ex}{\mbox{$\scriptstyle\ttype\toco\ttype$}}}
\newcommand{\nuiktwo}{\nu^\vari\kern-0.6ex\raisebox{-0.5ex}{\mbox{$\scriptstyle(\ttype\toco\ttype)\toco(\ttype\toco\ttype)$}}}

\newcommand{\vari}{\imath}
\newcommand{\varii}{\jmath}    

\newcommand{\cempty}{\mathord{\diamond}}

\newcommand{\kindpref}{kind-}
\newcommand{\rkindc}{\rulename{\kindpref{}c}}
\newcommand{\rkindvar}{\rulename{\kindpref{}var}}
\newcommand{\rkindabs}{\rulename{\kindpref{}abs}}
\newcommand{\rkindapp}{\rulename{\kindpref{}app}}

\newcommand{\leqpref}{leq$_{\mathsf{cf}}$-}

\newcommand{\rleqforalll}{\rulename{\leqpref{}\kern-0.2ex$\forall$\kern-0.3ex-l}}
\newcommand{\rleqforallr}{\rulename{\leqpref{}\kern-0.2ex$\forall$\kern-0.3ex-r}}
\newcommand{\rleqetal}{\rulename{\leqpref{}\kern-0.2ex$\eta$-l}}
\newcommand{\rleqetar}{\rulename{\leqpref{}\kern-0.2ex$\eta$-r}}
\newcommand{\rleqbetal}{\rulename{\leqpref{}\kern-0.2ex$\beta$-l}}
\newcommand{\rleqbetar}{\rulename{\leqpref{}\kern-0.2ex$\beta$-r}}
\newcommand{\rleqbeta}{\rulename{\leqpref{}\kern-0.2ex$\beta$}}

\newcommand{\rleqapp}{\rulename{\leqpref{}\kern-0.2ex app}}

\newcommand{\dleqpref}{leq-}

\newcommand{\rdleqsr}{\rulename{\dleqpref{}s-r}}
\newcommand{\rdleqinf}{\rulename{\dleqpref{}$\infty$}}

\newcommand{\rdleqforalll}{\rulename{\dleqpref{}\kern-0.2ex$\forall$\kern-0.3ex-l}}
\newcommand{\rdleqforallr}{\rulename{\dleqpref{}\kern-0.2ex$\forall$\kern-0.3ex-r}}
\newcommand{\rdleqbetal}{\rulename{\dleqpref{}\kern-0.2ex$\beta$-l}}
\newcommand{\rdleqbetar}{\rulename{\dleqpref{}\kern-0.2ex$\beta$-r}}
\newcommand{\rdleqbeta}{\rulename{\dleqpref{}\kern-0.2ex$\beta$}}
\newcommand{\rdleqlam}{\rulename{\dleqpref{}$\lambda$}}
\newcommand{\rdleqapp}{\rulename{\dleqpref{}\kern-0.2ex app}}
\newcommand{\rdleqappco}{\rulename{\dleqpref{}\kern-0.2ex app$\pco$}}
\newcommand{\rdleqappcontra}{\rulename{\dleqpref{}\kern-0.2ex app$\pcontra$}}
\newcommand{\rdleqapptop}{\rulename{\dleqpref{}\kern-0.2ex app$\ptop$}}
\newcommand{\rdleqrefl}{\rulename{\dleqpref{}refl}}
\newcommand{\rdleqasymm}{\rulename{\dleqpref{}antisym}}
\newcommand{\rdleqtrans}{\rulename{\dleqpref{}trans}}

\newcommand{\eq}{=}%
\newcommand{\eqpref}{eq-}
\newcommand{\reqbeta}{\rulename{\eqpref{}\kern-0.2ex$\beta$}}
\newcommand{\reqbetal}{\rulename{\eqpref{}\kern-0.2ex$\beta$-l}}
\newcommand{\reqbetar}{\rulename{\eqpref{}\kern-0.2ex$\beta$-r}}
\newcommand{\reqeta}{\rulename{\eqpref{}$\eta$}}
\newcommand{\reqinf}{\rulename{\eqpref{}$\infty$}}

\newcommand{\reqnab}{\rulename{\eqpref{}\kern-0.2ex$\nabla$}}

\newcommand{\reqc}{\rulename{\eqpref{}c}}
\newcommand{\reqvar}{\rulename{\eqpref{}var}}
\newcommand{\reqlam}{\rulename{\eqpref{}$\lambda$}}
\newcommand{\reqapp}{\rulename{\eqpref{}\kern-0.2ex app}}

\newcommand{\reqforall}{\rulename{\eqpref{}\kern-0.3ex$\forall$}}

\newcommand{\reqsym}{\rulename{\eqpref{}sym}}
\newcommand{\reqtrans}{\rulename{\eqpref{}trans}}

\newcommand{\cxtpref}{cxt-}
\newcommand{\rcxtempty}{\rulename{\cxtpref{}empty}}
\newcommand{\rcxtvar}{\rulename{\cxtpref{}var}}
\newcommand{\rcxttyvar}{\rulename{\cxtpref{}tyvar}}

\newcommand{\typref}{ty-}
\newcommand{\rtyc}{\rulename{\typref{}c}}
\newcommand{\rtyvar}{\rulename{\typref{}var}}
\newcommand{\rtyabs}{\rulename{\typref{}abs}}
\newcommand{\rtyapp}{\rulename{\typref{}app}}
\newcommand{\rtygen}{\rulename{\typref{}gen}}
\newcommand{\rtyinst}{\rulename{\typref{}inst}}
\newcommand{\rtysub}{\rulename{\typref{}sub}}
\newcommand{\rtyfold}{\rulename{\typref{}fold}}
\newcommand{\rtyunfold}{\rulename{\typref{}unfold}}

\newcommand{\rtyrec}{\rulename{\typref{}rec}}

\newcommand{\adm}[1]{\ #1\mbox{-}\tadm}
\newcommand{\tadm}{\mathsf{adm}}

\newcommand{\tshift}{\mathsf{shift}}
\newcommand{\tloop}{\mathsf{loop}}
\newcommand{\vloop}{\mathit{loop}}

\newcommand{\contpref}{cont-}

\newcommand{\rcontvar}{\rulename{\contpref{}var}}
\newcommand{\rcontabs}{\rulename{\contpref{}abs}}
\newcommand{\rcontapp}{\rulename{\contpref{}app}}
\newcommand{\rcontin}{\rulename{\contpref{}in}}
\newcommand{\rcontco}{\rulename{\contpref{}co}}
\newcommand{\rcontcontra}{\rulename{\contpref{}contra}}
\newcommand{\rcontsum}{\rulename{\contpref{}sum}}
\newcommand{\rcontprod}{\rulename{\contpref{}prod}}
\newcommand{\rcontarr}{\rulename{\contpref{}arr}}
\newcommand{\rcontforall}{\rulename{\contpref{}$\forall$}}
\newcommand{\rcontmu}{\rulename{\contpref{}mu}}
\newcommand{\rcontnu}{\rulename{\contpref{}nu}}

\newcommand{\ordpref}{ord-}
\newcommand{\rordvar}{\rulename{\ordpref{}var}}
\newcommand{\rords}{\rulename{\ordpref{}$\s$}}
\newcommand{\rordinf}{\rulename{\ordpref{}$\infty$}}
\newcommand{\jord}{\ \mathit{ord}}

\newcommand{\A}{\mathcal{A}}
\newcommand{\B}{\mathcal{B}}
\newcommand{\C}{\mathcal{C}}

\newcommand{\F}{\mathcal{F}}
\newcommand{\G}{\mathcal{G}}
\renewcommand{\H}{\mathcal{H}}
\renewcommand{\P}{\mathcal{P}}

\newcommand{\N}{\mathcal{N}}
\renewcommand{\S}{\mathcal{S}}

\newcommand{\calI}{\mathcal{I}}
\newcommand{\X}{\mathcal{X}}

\newcommand{\NAT}{\mathcal{N}\!\mathit{at}}
\newcommand{\LIST}{\mathcal{L}\!\mathit{ist}}
\newcommand{\STREAM}{\mathcal{S}\!\mathit{tream}}

\newcommand{\BOOL}{\mathcal{B}\mathit{ool}}

\newcommand{\LL}{\mathfrak{L}}

\newcommand{\rto}{\mathrel{{\raise.3ex\hbox{$\hookrightarrow$}}\kern-2.4ex{\lower.3ex\hbox{$\twoheadleftarrow$}}}}

\newcommand{\MU}{\boldsymbol{\mu}}
\newcommand{\NU}{\boldsymbol{\nu}}

\newcommand{\Inutwo}[2]{\NU^{#1}#2} 
\newcommand{\Imutwo}[2]{\MU^{#1}#2}

\newcommand{\subscriptopt}[1]{\ifthenelse{\equal{#1}{}}{}{_{#1}}}
\newcommand{\superscriptopt}[1]{\ifthenelse{\equal{#1}{}}{}{^{#1}}}

\newcommand{\lsub}[1][]{\sqsubseteq\superscriptopt{#1}}
\newcommand{\lSuper}[1][]{\sqsupseteq\superscriptopt{#1}}
\newcommand{\lsup}[1][]{\textstyle\bigsqcup\superscriptopt{#1}}
\newcommand{\linf}[1][]{\textstyle\bigsqcap\superscriptopt{#1}}

\newcommand{\lbot}[1][]{\textstyle\bot\superscriptopt{#1}}
\newcommand{\ltop}[1][]{\textstyle\top\superscriptopt{#1}}

\newcommand{\U}[1][]{\mathsf{U}\superscriptopt{#1}}

\newcommand{\tri}[1]{{^\rhd#1}}
\newcommand{\triA}{{^\rhd\!\!\A}}
\newcommand{\clos}{\overline}

\newcommand{\tin}[1][]{\mathsf{in}\superscriptopt{#1}}

\newcommand{\tout}[1][]{\mathsf{out}\superscriptopt{#1}}

\newcommand{\tfix}{\mathsf{fix}}
\newcommand{\tfixmu}{\tfix^\mu}\newcommand{\fixmu}{\tfixmu}
\newcommand{\tfixnu}{\tfix^\nu}\newcommand{\fixnu}{\tfixnu}
\newcommand{\tfixnabla}{\tfix^{\!\nabla}}
\newcommand{\nablakappa}{\nabla_{\!\kappa}}

\newcommand{\subst}[3]{[#1/#2]#3}
\newcommand{\update}[3]{#1[#2 \mapsto #3]}

\newcommand{\Nat}{\mathsf{Nat}}

\newcommand{\tzero}{\mathsf{zero}}
\newcommand{\tsucc}{\mathsf{succ}}

\newcommand{\tpred}{\mathsf{pred}}

\newcommand{\vas}{\mathit{as}}

\newcommand{\tcons}{\mathsf{cons}}

\newcommand{\Lam}{\mathsf{Lam}}

\newcommand{\tpair}{\mathsf{pair}}

\newcommand{\tfst}{\mathsf{fst}}
\newcommand{\tsnd}{\mathsf{snd}}

\renewcommand{\O}{\mathsf{O}}

\def\tovar#1{\setbox0=\hbox{$-^{#1}$}      
   \dimen0=\wd0                             
   \divide\dimen0 by 2                      %
   \mathrel{-\kern-0.42ex-^{\kern0.8ex\kern-\dimen0 #1}\kern-\dimen0\kern-0.2ex\longrightarrow}}

\newcommand{\Bool}{\mathsf{Bool}}
\newcommand{\ttrue}{\mathsf{true}}
\newcommand{\tfalse}{\mathsf{false}}
\newcommand{\Tree}{\mathsf{Tree}}

\newcommand{\V}{\mathsf{V}}

\newcommand{\tleaf}{\mathsf{leaf}}
\newcommand{\tnode}{\mathsf{node}}

\newcommand{\Rose}{\mathsf{Rose}}
\newcommand{\tstep}{\mathsf{step}}
\newcommand{\vstep}{\mathit{step}}
\newcommand{\tbf}{\mathsf{bf}}
\newcommand{\vbf}{\mathit{bf}}
\newcommand{\vrs}{\mathit{rs}}
\newcommand{\tappend}{\mathsf{append}}

\newcommand{\tmapStream}{\mathsf{mapStream}}

\newcommand{\tnats}{\mathsf{nats}}
\newcommand{\vnats}{\mathit{nats}}

\newcommand{\veq}{\mathit{eq}}

\newcommand{\Eq}{\mathsf{Eq}}

\newcommand{\uncount}{\beth}

\newcommand{\tmatchraw}{\mathsf{match}}
\newcommand{\twith}{\mathsf{with}}
\newcommand{\tmarr}{\mapsto}

\newcommand{\Int}{\mathop{\mathsf{Int}}\nolimits}

\newcommand{\PList}{\mathop{\mathsf{P\kern-0.2ex List}}\nolimits}
\newcommand{\PListF}{\mathop{\mathsf{P\kern-0.2ex ListF}}\nolimits}

\newcommand{\Bush}{\mathsf{Bush}}

\newcommand{\BTree}{\mathop{\mathsf{BTree}}\nolimits}

\newcommand{\LamE}{\mathop{\widehat{\mathsf{Lam}}}\nolimits}
\newcommand{\LamEF}{\mathop{\LamE\kern-0.4ex\mathsf{F}}\nolimits}
\newcommand{\tlame}{\mathop{\widehat{\mathsf{lam}}}\nolimits}
\newcommand{\tlamef}{\mathop{\tlame\kern-0.3ex\mathsf{f}}\nolimits}

\newcommand{\BT}{{\mathop{\mathsf{B\kern-0.2ex T}}\nolimits}}
\newcommand{\BTF}{{\mathop{\mathsf{B\kern-0.2ex TF}}\nolimits}}

\newcommand{\tappBT}{\mathop{\mathsf{appB\kern-0.2ex T}}\nolimits}
\newcommand{\tlamBT}{\mathop{\mathsf{lamB\kern-0.2ex T}}\nolimits}
\newcommand{\tapBT}{\mathop{\mathsf{apB\kern-0.2ex T}}\nolimits}
\newcommand{\tlmBT}{\mathop{\mathsf{lmB\kern-0.2ex T}}\nolimits}

\newcommand{\List}{\mathsf{List}}

\newcommand{\tnil}{\mathsf{nil}}

\newcommand{\Stream}{\mathsf{Stream}}

\newcommand{\Hungry}{\mathsf{Hungry}}
\newcommand{\tp}{\mathsf{p}}
\newcommand{\thu}{\mathsf{h}}
\newcommand{\ttr}{\mathsf{tr}}
\newcommand{\vtr}{\mathit{tr}}

\newenvironment{deffigure}[2]{%
  \def\deffigurecaption{#2}%
  \begin{figure*}[htbp]%
  \begin{center}%
  \begin{minipage}{#1}%
  \hrule \vspace*{4ex}%
}{%
\vspace{2ex} \hrule%
\addvspace{2ex}%
  \end{minipage}%
  \end{center}%
  \caption{\deffigurecaption}%
  \end{figure*}%
}

\renewcommand{\tfix}{\mathsf{fix}}

\renewcommand{\lv}{\llbracket}
\renewcommand{\rv}{\rrbracket}

\renewcommand{\V}{\mathcal{V}}

\newcommand{\NC}{\mathsf{NC}}
\newcommand{\NumClTree}{\mathsf{NumClTree}}
\renewcommand{\Fhat}{\ensuremath{\mathsf{F}_{\!\omega}\kern-0.4ex\widehat{}\kern0.40ex}}

\newcommand{\GRose}{\mathsf{GRose}}
\newcommand{\veqA}{\mathit{eqA}}
\newcommand{\veqF}{\mathit{eqF}}
\newcommand{\teqGRose}{\mathsf{eqGRose}}

\newcommand{\vfr}{\mathit{fr}}

\newcommand{\tloopnot}{\mathsf{loopnot}}
\newcommand{\vmaps}{\mathit{maps}}

\newcommand{\abbrfor}{\ \mbox{ for }\ }
\renewcommand{\adm}[1]{\ {#1}\mbox{-}\tadm}
\newcommand{\sapp}{\cdot}
\newcommand{\tss}{\mathsf{s}}
\renewcommand{\th}{\mathsf{h}}

\renewcommand{\lsup}[1][]{\sup\nolimits\superscriptopt{#1}}
\renewcommand{\linf}[1][]{\inf\nolimits\superscriptopt{#1}}

\newenvironment{figone}[2]{%
  \def\deffigurecaption{#2}%
  \begin{figure}[htbp]%
  \begin{center}%
  \begin{minipage}{#1}%
  \hrule \vspace*{4ex}%
}{%
\vspace{2ex} \hrule%
\addvspace{2ex}%
  \end{minipage}%
  \end{center}%
  \caption{\deffigurecaption}%
  \end{figure}%
}

\newcommand{\sep}{;}

\def\doi{4 (2:3) 2008}
\lmcsheading%
{\doi}
{1--33}
{}
{}
{Jan.~\phantom{0}4, 2007}
{Apr.~10, 2008}
{}

\begin{document}

\title{Semi-continuous Sized Types and Termination\rsuper *}
\author[A.~Abel]{Andreas Abel}
\address{
Institut f\"ur Informatik \\
Ludwig-Maximilians-Universit\"at M\"unchen
}
\email{abel@tcs.ifi.lmu.de}
\thanks{
    Research supported by the coordination action \emph{TYPES}
    (510996) and  thematic network \emph{Applied Semantics II} (IST-2001-38957)
    of the European Union and the project \emph{Cover} of the Swedish
    Foundation of Strategic Research (SSF)}
\titlecomment{{\lsuper *}A shorter version of this article has appeared in the
  proceedings of Computer Science Logic 2006 \cite{abel:csl06}.}

\begin{abstract}
  Some type-based approaches to termination use sized types:
  an ordinal bound for the size of a data structure is stored in its
  type.  A recursive function over a sized type is accepted if it
  is visible in the type system that recursive calls occur just at a
  smaller size.  This approach is only sound if the type of the
  recursive function is admissible, i.e., depends on the size index in
  a certain way.  To explore the space of admissible functions in the
  presence of higher-kinded data types and impredicative polymorphism, a
  semantics is developed where sized types are interpreted as
  functions from ordinals into sets of strongly normalizing terms.  It
  is shown that upper semi-continuity of such functions is a
  sufficient semantic criterion for admissibility.  To provide a
  syntactical criterion, a calculus for semi-continuous functions is developed.
\end{abstract}

\maketitle

\section{Introduction}
\label{sec:intro}

Termination of computer programs has received continuous interest in
the history of computer science, and classical applications are total
correctness and termination of partial evaluation.  In languages with
a notion of computation on the type-level, such as dependently-typed
languages or rich typed intermediate languages in compilers
\cite{crary:lx}, termination of expressions that compute a type is
required for type checking and type soundness.  Further, theorem
provers that are based on the Curry-Howard Isomorphism and offer a
functional programming language to write down proofs
usually reject non-terminating programs to ensure consistency.  
Since the
pioneering work of Mendler \cite{mendler:lics}, termination analysis
has been combined with typing,
with much success for strongly-typed languages \cite{pareto:sizedtypes,amadio:guardcondition,gimenez:strec,xi:termination,gimenez:typeBased,blanqui:rta04}.  The resulting
technique, \emph{type-based termination checking}, has several
advantages over a purely syntactical termination analysis: (1) It is
\emph{robust} \wrt\ small changes of the analyzed program, since it is
working on an abstraction of the program: its type.  So if the
reformulation of a program (e.g., by introducing a redex) still can be
assigned the same sized type, it automatically passes the termination
check.  (2) In design and justification, type-based termination rests
on a technology extensively studied for several decades: types.  (3)
Type-based termination is essentially a refinement of the typing rules
for recursion and for introduction and elimination of data.  This is
\emph{orthogonal} to other language constructs, like variants, records, and
modules.  Thus, a language can be easily enriched by such constructs without change to
the termination checker.  This is not true if termination checking is a
separate static analysis.  Orthogonality has an especially pleasing
effect: (4) Type-based termination scales to \emph{higher-order functions}
and \emph{polymorphism}.  (5) Last but not least, it effortlessly creates a
termination \emph{certificate}, which is just the typing derivation.

Type-based termination especially plays its strength when combined
with higher-order datatypes and higher-rank polymorphism, \ie,
occurrence of $\forall$ to the left of an arrow.  Let us see
an example.  We consider the type of generalized rose trees $\GRose\,F A$
parameterized by an element type $A$ and the branching type $F$.
It is given by two constructors:
\[
\begin{array}{lll}
\tleaf  & : & \GRose\,F A \\
\tnode & : & A \to F\,(\GRose\,F A) \to \GRose\,F A
\end{array}
\]
Generalized rose trees are either a $\tleaf$ or a 
$\tnode\,a\,\vfr$ of a label $a$ of type $A$ and a collection of subtrees
$\vfr$ of type $F\,(\GRose\,F A)$.  Instances of generalized rose
trees are binary trees ($F A = A \times A$), finitely branching trees
($F A = \List\,A$), or infinitely branching trees ($F A = \Nat \to A$).
Programming a
generic equality function for generalized rose trees that is polymorphic in $F$
and $A$, we will end up with the following equations:
\[
\begin{array}{l}
  \Eq\,A  =  A \to A \to \Bool 
\\[1.5ex]
  \teqGRose  :  (\forall A.\, \Eq\,A \to \Eq\,(F A)) \to 
    \forall A.\, \Eq\,A \to \Eq\,(\GRose\,F A)
\\[1ex]
  \teqGRose\ \veqF\,\veqA\ \tleaf\ \tleaf = \ttrue \\
  \teqGRose\ \veqF\,\veqA\ (\tnode\;a\;\vfr)\ (\tnode\;a'\;\vfr') = 
  \begin{array}[t]{l}
 (\veqA\;a\;a') \wedge {}\\
 (\veqF\;(\teqGRose\;\veqF\,\veqA)\;\vfr\;\vfr') \\
  \end{array}\\
  \teqGRose\ \veqF\,\veqA\ \_\ \_ = \tfalse \\
\end{array}
\]
The generic equality $\teqGRose$ takes two parametric arguments,
$\veqF$ and $\veqA$.  The second one is a placeholder for an
equality test for type $A$, the first one lifts an equality test
for an arbitrary type $A$ to an equality test for the 
type $F A$.  The equality test for generalized rose trees, 
$\teqGRose\;\veqF\,\veqA$, is then defined by recursion on the next two
arguments.  In the case of two $\tnode$s we would expect a
recursive call, but instead, the function itself is passed as an
argument to $\veqF$, one of its own arguments!  Nevertheless,
$\teqGRose$ is a total function, provided its arguments are total
and well-typed.  However, with traditional methods, which only take
the computational behavior into account, it will be hard to verify
termination of $\teqGRose$.  This is due to the fact that the
polymorphic nature of $\veqF$ plays a crucial role.  It is easy to
find an instance of $\veqF$ of the wrong type which makes the program
loop.  Take, for instance:
\[
\begin{array}{l}
  \veqF : \Eq\,(\GRose\,F\,\Nat) \to \Eq\,(F\,(\GRose\,F\,\Nat)) \\
  \veqF\,\veq\,\vfr\,\vfr' =
    \veq\,(\tnode\,0\,\vfr)\,(\tnode\,0\,\vfr') \\
\end{array}
\]

A type-based termination criterion however passes $\teqGRose$ with
ease:  Consider the indexed type $\GRose^\vari\,F A$ of generalized rose trees
whose height is smaller than $\vari$.  The types of the constructors are
refined as follows:
\[
\begin{array}{lll}
  \tleaf & : & \forall F \forall A\forall\vari.\,
    \GRose^{\vari+1}\,F A \\
  \tnode & : & \forall F \forall A\forall\vari.\ A \to
    \GRose^\vari\,F A \to 
    \GRose^{\vari+1}\,F A \\
\end{array}
\]
When defining $\teqGRose$ for
trees of height $< \vari+1$, we may use $\teqGRose$ on trees of height
$<\vari$.  Hence, in the clause for two $\tnode$s, term
$\teqGRose\;\veqF\;\veqA$
has type $\Eq\,(\GRose^\vari\,F A)$, and $\veqF\;(\teqGRose\;\veqF\;\veqA)$
gets type $\Eq\,(F\,(\GRose^\vari\,F A))$, by instantiation of the
polymorphic type of $\veqF$.  Now it is safe to apply the last
expression to $\vfr$ and $\vfr'$ which are in $F\,(\GRose^\vari\,F A)$,
since $\tnode\,a\,\vfr$ and $\tnode\,a'\,\vfr'$ were assumed to be
in $\GRose^{\vari+1}\,F A$.

In essence, type-based termination is a stricter typing of the
fixed-point combinator $\tfix$ which introduces recursion.  The
unrestricted use, via the typing rule (1),
is replaced by a rule with a stronger hypothesis (2):
\begin{gather*}
  \mbox{(1)}~~ \ru{f : A \to A}{\tfix\,f : A}
\qquad\qquad
  \mbox{(2)}~~ \ru{f : \forall \vari.\,A(\vari) \to A(\vari+1)
        }{\tfix\,f : \forall n.\, A(n)}
\end{gather*}
Soundness of rule (2) can be shown by induction on $n$.  To get
started, we need to show $\tfix\,f : A(0)$ which requires $A(\vari)$ to be
of a special shape, for instance $A(\vari) = \GRose^\vari\,F\,B \to C$
(this corresponds to Hughes, Pareto, and Sabry's \emph{bottom check} \cite{pareto:sizedtypes}).
Then $A(0)$ denotes functions which have to behave well for
all arguments in $\GRose^0\,F\,B$, \ie, for no arguments, since
$\GRose^0\,F\,B$ is empty.  Trivially, any program fulfills this
condition.   In the step case, we need to show $\tfix\,f : A(n+1)$,
but this follows from the equation $\tfix\,f = f\,(\tfix\,f)$ since $f
: A(n) \to A(n+1)$, and $\tfix\,f : A(n)$ by induction hypothesis.

In general, the index $\vari$ in $A(\vari)$ will be an \emph{ordinal} number.
Ordinals are useful when we want to speak of objects of unbounded size,
\eg, generalized rose trees of height $< \omega$ that
inhabit the type $\GRose^\omega\,F A$.  Even more, ordinals are
required to denote the height of infinitely branching trees: take generalized rose
trees with $F A = \Nat \to A$.  
Other examples of infinite branching, which come from the area of
type-theoretic theorem provers, are the $W$-type,
Brouwer ordinals and the accessibility predicate \cite{paulin:inductiveTR}.

In the presence of ordinal indices, rule (2) has to be
proven sound by transfinite induction.  In the case of a limit ordinal
$\lambda$, we have to infer $\tfix\,f : A(\lambda)$ from the
induction hypothesis $\tfix\,f : \forall \alpha<\lambda.\,
A(\alpha)$.  This imposes extra conditions on the shape of a so-called
\emph{admissible} type $A$, which
are the object of this article.  Of course, a monotone $A$ is
trivially admissible, but many interesting types for recursive
functions are not monotone, like $A(\alpha) = \Nat^\alpha \to \Nat^\alpha \to
\Nat^\alpha$ (where $\Nat^\alpha$ contains the natural numbers $<\alpha$).
We will show that all those
types $A(\alpha)$ are admissible  that are \emph{upper semi-continuous} in $\alpha$,
meaning $\limsup_{\alpha\to\lambda} \A(\alpha) \subseteq \A(\lambda)$
for limit ordinals $\lambda$.  
Function types $C(\alpha) = A(\alpha) \to B(\alpha)$
will be admissible if $A$ is \emph{lower semi-continuous} ($A(\lambda)
\subseteq \liminf_{\alpha\to\lambda}\A(\alpha)$) and $B$ is
upper semi-continuous.  Similar laws will be developed for the other
type constructors and put into the form of a kinding system for
semi-continuous types.

Before we dive into the mathematics, let us make sure that
semi-continuity is relevant for termination.  A type which
is not upper semi-continuous is $A(\vari) = (\Nat^\omega \to
\Nat^\vari) \to \Nat^\omega$ (see Sect.~\ref{sec:negres}).  
Assuming we can nevertheless use this
type for a recursive function, we can construct a loop.  
First, define
successor $\tsucc : \forall \vari.\, \Nat^\vari \to \Nat^{\vari+1}$ and predecessor $\tpred: \forall \vari.\, \Nat^{\vari+1} \to \Nat^\vari$.
Note that the size index is an upper bound and $\omega$ is the biggest
such bound for the case of natural numbers, thus, we have the
subtyping relations $\Nat^\vari \leq \Nat^{\vari+1} \leq \dots \leq
\Nat^\omega \leq \Nat^{\omega+1} \leq \Nat^\omega$.

We make the following definitions:
\[
\begin{array}[t]{lll}
 A(\vari) & := &  (\Nat^\omega \to \Nat^\vari) \to \Nat^\omega \\[2ex]
\tshift & : & \forall \vari.\, (\Nat^\omega \to \Nat^{\vari+1}) \\&& \bla 
  \to \Nat^\omega \to \Nat^\vari \\
\tshift & := & \lambda g \lambda n.\, \tpred\,(g\,(\tsucc\,n)) 
\end{array}
\qquad\qquad
\begin{array}[t]{lll}
 f & :  & \forall\vari.\, A(\vari) \to A(\vari+1) \\
 f & := & \lambda\vloop\lambda g.\  
          \vloop\,(\tshift\,g)
\\[2ex]
 \tloop & : & \forall\vari.\, A(\vari) \\
 \tloop & := & \tfix\,f
\end{array}
\]
Since $\Nat^\omega \to \Nat^0$ is empty, $A$ passes the bottom check. 
Still, instantiating types to $\tsucc : \Nat^\omega \to \Nat^\omega$ and
$\tloop : (\Nat^\omega \to \Nat^\omega) \to \Nat^\omega$ we convince
ourselves that the execution of $\tloop\;\tsucc$ indeed runs forever.

\subsection{Related Work and Contribution}

Ensuring termination through typing is quite an old idea, just think
of type systems for the $\lambda$-calculus like simple types, System
$\Fpure$, System $\Fomega$, or the Calculus of Constructions, which all
have the normalization property.  These systems have been extended by
special recursion operators, like primitive recursion in G\"odel's T,
or the recursors generated for inductive definitions in Type Theory
(\eg, in Coq%
). These recursion operators preserve normalization but limit the
definition of recursive functions to special patterns, namely
instantiations of the recursion scheme dictated by the recursion
operator.  Taming general recursion  $\tfix\,f$ through
typing, however, which allows the definition of recursive functions in
the intuitive way known from functional programming, is not yet fully
explored.  Mendler \cite{mendler:lics} pioneered this field; he used a
certain polymorphic typing of the functional $f$ to obtain primitive
(co)recursive functions over arbitrary datatypes.  Amadio and
Coupet-Grimal \cite{amadio:guardcondition} and \gim\
\cite{gimenez:strec} developed Mendler's approach further, until 
a presentation using ordinal-indexed (co)inductive types was found and
proven sound by Barthe \etal\ \cite{gimenez:typeBased}.  The system
$\lambdahat$ presented in \loccit{} restricts types $A(\vari)$ of
recursive functions to the shape $\mu^\vari F \to C(\vari)$ where the
domain must be an inductive type $\mu^\vari F$ indexed by $\vari$ and
the codomain a type $C(\vari)$ that is monotonic in $\vari$.  This
criterion, which has also been described by the author
\cite{abel:rairo04}, 
allows for a simple soundness proof in the limit case of the
transfinite induction, but excludes interesting types like the considered
\[
  \Eq\,(\GRose^\vari\,F A) = \GRose^\vari\,F A \to
  \GRose^\vari\,F A \to \Bool
\]
which has an antitonic codomain $C(\vari) = \GRose^\vari\,F A \to
\Bool$.  The author has in previous work widened the criterion, but
only for a type system without polymorphism \cite{abel:tlca03}.  Other
recent works on type-based termination
\cite{blanqui:rta04,blanqui:csl05,bartheGregoirePastawski:tlca05} stick
to the restriction of $\lambdahat$.  Xi \cite{xi:termination} uses
dependent types and lexicographic measures to ensure termination of
recursive programs in a call-by-value language, but his indices are
natural numbers instead of ordinals; this excludes infinite objects we
are interested in.  

Closest to the present work is the sized type system of Hughes,
Pareto, and Sabry \cite{pareto:sizedtypes}, \emph{Synchronous Haskell}
\cite{pareto:PhD}, which admits ordinal indices up to $\omega$.  Index
quantifiers as in $\forall \vari.\, A(\vari)$ range over natural
numbers, but can be instantiated to $\omega$ if $A(\vari)$ is
\emph{$\omega$-undershooting}.  Sound semantic criteria for
$\omega$-undershooting types are already present, but in a somewhat
ad-hoc manner.  We cast these criteria in the established mathematical
framework of semi-continuous functions and provide a syntactical
implementation in form of a derivation system.  Furthermore, we
allow ordinals
beyond $\omega$ 
and infinitely branching inductive types that
invalidate some criteria for the only finitely branching
tree types in \emph{Synchronous Haskell}.  Finally, we allow
polymorphic recursion, 
impredicative polymorphism and higher-kinded inductive and
coinductive types such as $\GRose$.
This article summarizes the main results of the author's dissertation
\cite{abel:PhD}.  A shorter version has appeared in the CSL'06
proceedings \cite{abel:csl06}.

\subsection{Contents}

In Section~\ref{sec:syntax} we introduce the syntax of $\Fhat$, our
$\lambda$-calculus with higher-kinded polymorphism, recursion over
higher-kinded inductive types and corecursion into higher-kinded
coinductive types.  Static semantics (\ie, typing rules) and dynamic
semantics (\ie, reduction rules) are presented there, and we formally
express the $\teqGRose$-example from the introduction in $\Fhat$.  In
Section~\ref{sec:sem} we model the types of $\Fhat$ as saturated sets
of strongly normalizing terms in order to show termination of
well-typed programs.  After these two technical sections we come to
the main part of this article:  In Section~\ref{sec:semicont} we identify
compositional criteria for semi-continuous types and in
Section~\ref{sec:negres} we justify the absence of certain composition
schemes by giving counterexamples.  These results are put in the form of a
calculus for semi-continuous types in Section~\ref{sec:der},
culminating in syntactic rules for admissible (co)recursion types.  We
close by giving some practical examples for admissible types.

\subsection{Preliminaries}

We assume that the reader is to some extent acquainted with the higher-order
polymorphic lambda-calculus, System~\Fomega\ (see Pierce's text book
\cite{pierce:tapl}) and has some knowledge of ordinals, inductive types, and
strong normalization. 

\begin{deffigure}{\textwidth}{\Fhat: Syntax and operational semantics.\label{fig:syntax}}
Polarities, kinds, constructors, kinding contexts.
\[
\def\arraystretch{1.2}
\begin{array}{lrl@{\hspace{4ex}}l}
  p & ::= & \pco \mid \pcontra \mid \pin 
    & \mbox{polarity} \\
  \kappa & ::= & \ttype \mid \tord \mid p\kappa \to \kappa' 
    & \mbox{kind} \\
  \pkind & ::= & \ttype \mid p\pkind \to \pkind'
    & \mbox{pure kind} \\
  a,b,A,B,F,G & ::= & C \mid X \mid \lambda X\ofkinddot\kappa F \mid F\,G
    & \mbox{(type) constructor} \\
  C & ::= & 1 \mid + \mid \times \mid {\to} \mid \forall_\kappa \mid
    \muka \mid \nuka \mid \s \mid \infty
    & \mbox{constructor constants} \\
  \Delta & ::= & \cempty \mid \Delta, X\of p\kappa
    & \mbox{kinding context}
\end{array}
\]
Constructor constants and their kinds  ($\kappa \tox p
\kappa'$ means $p\kappa \to \kappa'$).
\[
\def\arraystretch{1.2}
\begin{array}[t]{lll@{\hspace{4ex}}l}
 1      & : & \ttype
    & \mbox{ unit type } \\
 +      & : & \ttype \toco \ttype \toco \ttype
    & \mbox{ disjoint sum }  \\
 \times & : & \ttype \toco \ttype \toco \ttype
    & \mbox{ cartesian product }  \\
 \mbox{$\to$} & : & \tominus \ttype {\toplus \ttype \ttype}
    & \mbox{ function space } \\
 \forall_\kappa & : & \toplus{(\tozero \kappa \ttype)}\ttype
    & \mbox{ quantification }\\
 \muka & : & \toplus \tord \kfix 
    & \mbox{ inductive constructors} \\
 \nuka & : & \tominus \tord \kfix
    & \mbox{ coinductive constructors} \\
 \s & : & \toplus \tord \tord 
    & \mbox{ successor of ordinal } \\
 \infty & : & \tord
    & \mbox{ infinity ordinal } \\
\end{array}
\]
Objects (terms), values, evaluation frames, typing contexts.
\[
\def\arraystretch{1.2}
\begin{array}{lrl@{\hspace{4ex}}l}
  r, s, t \!\!&\!\! ::= &\!\! c \mid x \mid \lambda x t \mid r\,s 
  &\!\! \mbox{term} 
\\
  c \!\!&\!\! ::= &\!\! \ttuple{} \mid \tpair \mid \tfst \mid \tsnd
    \mid \tinl \mid \tinr \mid \tcaseraw \mid \tin \mid \tout
    \mid \tfixmu_n \mid \tfixnu_n 
  &\!\! \mbox{constant ($n\in\NN$)}
\\
  v \!\!&\!\! ::= &\!\! \lambda x t \mid \tpair\,t_1\,t_2 \mid \tinl\,t
                     \mid \tinr\,t \mid \tin\,t \mid c \mid \tpair\,t 
             \mid \tfixnabla_n s\,t_{1..m} 
    &\!\! \mbox{value}\mbox{ ($m \leq n$) }
\\
  e(\_) \!\!&\!\! ::= &\!\! \_\,s \mid \tfst\,\_ \mid \tsnd\,\_ \mid \tcaseraw\,\_
                \mid \tout\,\_ \mid \tfixmu_n\, s\ t_{1..n}\, \_
    &\!\! \mbox{evaluation frame} \\
  E(\_) \!\!&\!\! ::= &\!\! e_1(\dots e_n(\_)\dots)
    &\!\! \mbox{evaluation context ($n \geq 0$)} \\
  \Gamma \!\!&\!\! ::= &\!\! \cempty \mid \Gamma, x\of A \mid \Gamma,
    X\ofkind{p\kappa}
  &\!\! \mbox{typing context}
\\
\end{array}
\]
Reduction $t \red t'$.
\[
\begin{array}[t]{lll@{\hspace{4ex}}l}
  (\lambda x t)\, s & \red & \subst s x t \\
  \tfst\,\ttuple{r,s} & \red & r \\
  \tsnd\,\ttuple{r,s} & \red & s \\
  \tcaseraw\,(\tinl\,r) & \red & \lambda x\lambda y.\,x\,r \quad (*) \\
  \tcaseraw\,(\tinr\,r) & \red & \lambda x\lambda y.\,y\,r \quad (*) \\
\end{array}
\qquad
\begin{array}[t]{lll@{\hspace{4ex}}l}
  \tout\,(\tin\,r) & \red & r \\
  \tfixmu_n\, s\ t_{1..n}\, (\tin\,t) & \red & 
    s\, (\tfixmu_n\, s)\, t_{1..n}\, (\tin\,t) \\
  \tout\,(\tfixnu_n\, s\, t_{1..n}) & \red & 
    \tout\,(s\, (\tfixnu_n\, s)\, t_{1..n})
  \\
  \\
  \multicolumn 3 l {\mbox{+ closure under all term constructs}}
\end{array}
\]
(*) $x,y \not\in\FV(r)$.
\end{deffigure}

\section{Overview of System \Fhat}
\label{sec:syntax}

In this section we introduce $\Fhat$, an \emph{a posteriori} strongly
normalizing extension of System $\Fomega$ with higher-kinded inductive
and coinductive types and (co)recursion combinators.
Figure~\ref{fig:syntax} summarizes the syntactic entities.

\para{Type constructors}
We seek to model sized types like $\GRose^\vari F\,A$ whose first
parameter $F$ is a type \emph{constructor} of kind $\ttype \to
\ttype$, meaning that it maps types to types.  It is therefore
suggestive to take $\Fomega$ as basis, which formalizes type
constructors of arbitrary kind and, \eg, lays the foundation for the purely
functional language Haskell.  In the introduction, we have presented
$\GRose$s as built from two (data) constructors $\tleaf$ and $\tnode$;
however, for a theoretic analysis it is more convenient to consider
$\GRose\,F\,A$ as the least fixed-point of the type constructor
$\lambda X.\, 1 + (A \times F\,X)$.  For this we write
\[
  \GRose\,F A := \mu \lambda X.\, 1 + (A \times F\,X)
  .
\]
Herein, $1$ is the unit type and $+$ the disjoint sum.  Taking the
empty tuple $\ttuple{} : 1$ to be the inhabitant of the unit type and $\tinl
: A \to (A + B)$ and $\tinr : B \to (A + B)$ the two injections into
the disjoint sum lets us \emph{define} the original data constructors:
\[
\begin{array}{lll}
  \tleaf  & : & \GRose\,F A \\
  \tleaf & := &  \tin\,(\tinl\, \ttuple{})
\\[1ex]
  \tnode & : & A \to F\,(\GRose\,F A) \to \GRose\,F A \\
  \tnode & := & \lambda a \lambda \vfr.\, \tin\,(\tinr\, \ttuple{a,\vfr})
\end{array}
\]
(The tag $\tin$ introduces a inductive type, see below.)

\para{Polarized kinds}
Negative recursive types such as $\mu \lambda X.\,X \to 1$ allow the
coding of $\mathsf{Y}$ and other fixed-point combinators as pure
$\lambda$-terms, so one can write recursive programs without special
syntax for recursion \cite{mendler:lics}.  For our purposes, this is
counter-productive---type systems for termination need to identify all
uses of recursion.  Therefore, we restrict to positive recursive types
$\mu H$ where $H$ is monotone.  In the case of $\GRose$, the
underlying constructor $H\,X = 1 + (A \times F\,X)$ must be monotone,
which is the case if $F$ is monotone.  So $\GRose\,F A$ is only
well-formed for monotone $F$.  To distinguish type constructors by
their monotonicity behavior, also called \emph{variance}, 
we equip function kinds  with polarities $p$
\cite{steffen:PhD}, which are written before the domain or on top of
the arrow.  Polarity $\pco$ denotes covariant constructors, $\pcontra$
contravariant constructors and $\pin$ mixed-variant constructors
\cite{dugganCompagnoni:subtyping}.  For instance:
\[
\begin{array}{lll}
  \lambda X.\, X \to 1 & : & \ttype \tocontra \ttype \\
  \lambda X.\, X \to X & : & \ttype \toin \ttype \\
  \lambda X.\, \Int \to (1 + X) & : & \ttype \toco \ttype \\
  \GRose & : & (\ttype \toco \ttype) \toco \ttype \toco \ttype \\
\end{array}
\]
Abel \cite{abel:csr06} and Matthes \cite{abelMatthes:csl04} provide
more explanation on polarities.

\para{Sized inductive types}
We refine inductive types $\mu F$ to sized inductive types $\mu^a F$. 
The first argument, $a$, to $\mu$, which we
usually write as superscript, denotes the upper bound for the height
of data represented by terms of the inductive type.  
The index $a$ is a constructor of
kind $\tord$ and denotes an ordinal; the relevant ordinal expressions
are given by the grammar
\[
  a ::= \vari \mid \s\,a \mid \infty
\]
with $\vari$ an ordinal variable.\footnote{One could add a constant
  for the ordinal $0$, but for our purposes it is enough that each
  concrete data structure inhabits $\mu^\infty F$. For checking
  termination relative sizes are sufficient, which can be expressed
  using ordinal variables and successor.}
If $a$ actually denotes a finite
ordinal (a natural number), then the height is simply the number of
data constructors on the longest path in the tree structure of any element
of $\mu^a F$.  Since $a$ is only an upper bound, $\mu^a F$ is a
subtype of $\mu^b F$, written $\mu^a F \leq \mu^b F$ for $a \leq b$, 
meaning that $\mu$ is covariant in the
index argument.  Finally, $F \leq F'$ implies $\mu^a F \leq \mu^a F'$,
so we get the kinding
\[
  \mu : \tord \toco (\ttype \toco \ttype) \toco \ttype
\]
for the least fixed-point constructor. For the \emph{closure ordinal} $\infty$,  we have
\[
  \mu^\infty F = \mu^{\infty+1} F ,
\]
where $\infty+1$ is a shorthand for $\s\infty$, $\s:\tord \toco \tord$
being the successor on ordinals.  

Because $\infty$ denotes the closure ordinal, the
axiom $\s\,\infty = \infty$ is justified.  Equality on type constructors
is defined as the least congruent equivalence relation closed under
this equation and $\beta\eta$.

At this point, let us stress that
the \emph{syntax} of ordinals is extremely simple, hence, equality of
types and subtyping is decidable.  The user can think of ordinals as
of natural numbers with infinity, although they will be interpreted as
real ordinals up to a fairly large closure ordinal in Section~\ref{sec:sem}.

\begin{exa}[Some sized types]
\[
\begin{array}[t]{lll}
  \Nat  & :  & \tord \toco \ttype \\
  \Nat  & := & \lambda \vari.\ \mu^\vari \lambda X.\, 1 + X 
\\[1ex]
  \List & :  & \tord \toco \ttype \toco \ttype \\
  \List & := & \lambda \vari\lambda A.\ \mu^\vari \lambda X.\, 1 + A
    \times X 
\end{array}
\quad
\begin{array}[t]{lll}
  \GRose & :  & \tord \toco (\ttype \toco \ttype) \toco \ttype \toco \ttype \\
  \GRose & := & \lambda \vari\lambda F\lambda A.\ 
                  \mu^\vari \lambda X.\, 1 + A \times F\,X 
\\[1ex]
  \Tree & :  & \tord \toco \ttype \tocontra \ttype \toco \ttype \\
  \Tree & := & \lambda \vari \lambda B \lambda A.\,   
                   \GRose^\vari\,(\lambda X.\, B \to X)\,A
\end{array}
\]  
\end{exa}

\para{Sized coinductive types}
Dually to inductive or least fixed-point types $\mu F$ we have
coinductive or greatest fixed-point types $\nu F$ to model infinite
structures.  For instance $\Stream A = \nu X.\, A \times X$ contains
the infinite sequences over $A$.  The dual to the height of an
inductive data structure is the \emph{depth} of a coinductive
one, i.e., how often one can unwind the structure.  So the size $a$ of
a sized coinductive type $\nu^a F$ is a lower bound on the depth of
its inhabitants.  Since it is a lower bound, coinductive types are
contravariant in their size index:
\[
\nu : \tord \tocontra (\ttype \toco \ttype) \toco \ttype
.
\]
As for inductive types, the equation $\nu^\infty F = \nu^{\infty + 1}
F$ holds.
\begin{exa}[Sized streams]
On a stream in $\Stream^a A$ one can safely read off the first
$a$ elements. 
\[
\begin{array}{lll}
  \Stream & :  & \tord \tocontra \ttype \toco \ttype \\
  \Stream & := & \lambda \vari\lambda A.\, \nu^\vari \lambda X.\, A \times X 
\end{array}
\]  
\end{exa}

\para{Heterogeneous datatypes}
If we consider not only fixed-point \emph{types}, but also fixed-point
\emph{constructors}, we can treat programs involving so-called nested
or heterogeneous types.  A simple example of a heterogeneous type is
the type of powerlists $\PList A$ which contains lists of $A$s whose
length is a power of two \cite{hinze:efficientGfolds}.  
The type constructor $\PList : \ttype \toco \ttype$ can be
modeled as $\mu \lambda X \lambda A.\, A + X\,(A \times A)$ which is
the least fixed-point of a type constructor of kind $(\ttype \toco
\ttype) \toco (\ttype \toco \ttype)$.

Sized heterogeneous types are obtained by simply generalizing $\mu$
and $\nu$ to
\[
\begin{array}{lll}
  \mu_\kappa & : & \tord \toco (\kappa \toco \kappa) \toco \kappa
\\
  \nu_\kappa & : & \tord \tocontra (\kappa \toco \kappa) \toco \kappa
.
\end{array}
\]
The kind $\kappa$ is required
to be \emph{pure}, \ie, a kind not mentioning $\tord$, for reasons
explained in Section~\ref{sec:closord}.  All our examples work for
pure $\kappa$.

\begin{exa}[Sized heterogeneous types]
\[
  \begin{array}{lll}
    \PList & :  & \tord \toco \ttype \toco \ttype \\
    \PList & := & \lambda \vari.\ \mu^\vari \lambda X \lambda A.\, A + X\,(A \times A)
    \\[1.5ex]
    \Bush  & :  & \tord \toco \ttype \toco \ttype \\
    \Bush  & := & \lambda \vari.\ \mu^\vari \lambda X \lambda A.\, 1 + A \times
    X\,(X\, A) \\[1.5ex]
    \Lam   & :  & \tord \toco \ttype \toco \ttype \\
    \Lam   & := & \lambda \vari.\ \mu^\vari \lambda X \lambda A. \, A + X\,A \times
    X\,A + X\,(1 + A) \\[1.5ex]
  \end{array}
\]  
The second type,
$\Bush^a\,A$, bushy lists, models finite maps from unlabeled binary
trees of height $< a$ into $A$
\cite{alti:coinductiveRep,hinze:GGTries}.  The third
type, $\Lam^a\,A$, is inhabited by de Bruijn representations of
untyped lambda terms of height $< a$ with free variables in $A$
\cite{bird:debruijn,alti:monadic}. 
\end{exa}

\para{Programs}
The term language of $\Fhat$ is the $\lambda$-calculus plus the standard
constants to
introduce and eliminate unit ($1$), sum ($+$), and product ($\times$)
types.  We write $\ttuple{t_1,t_2}$ for $\tpair\,t_1\,t_2$.
Further, there is folding, $\tin$, and unfolding, $\tout$,
of (co)inductive types.  The complete listing of the typing rules can
be found in Figure~\ref{fig:termty} in the appendix, here we discuss the
most important ones.
Let $\kappa = \vec p\vec\kappa \to \ttype$ a
pure kind, $F : \pco\kappa \to \kappa$, $G_i : \kappa_i$ for $1 \leq i
\leq |\vec\kappa|$,  $a : \tord$, and $\nabla \in \{ \mu,\nu\}$,
then we have the following (un)folding rules:
\begin{gather*}
  \nru{\rtyfold
     }{\Gamma \der t : F\, (\nablakappa^a\, F)\, \vec G
     }{\Gamma \der \tin\,t : \nablakappa^{a+1} F\, \vec G}
\qquad
  \nru{\rtyunfold
     }{\Gamma \der r :  \nablakappa^{a+1} F\, \vec G
     }{\Gamma \der \tout\,r : F\, (\nablakappa^a\, F)\, \vec G}
\end{gather*}
Finally, there are fixed-point combinators $\tfixmu_n$ and $\fixnu_n$
for each $n \in \NN$ on the term level.  The term $\tfixmu_n\,s$
denotes a recursive function with $n$ leading non-recursive arguments;
the $n+1$st argument must be of an inductive type.  Similarly,
$\tfixnu_n\,s$ is a corecursive function which takes $n$ arguments and
produces an inhabitant of a coinductive type.  
We abbreviate $f\,t_1\,\dots\,t_n$ by $f\,t_{1..n}$ or $f\,\vec t$.

One-step reduction $t \red t'$ is defined by the $\beta$-reduction
axioms given in Figure~\ref{fig:syntax} plus congruence
rules.  
\longversion{Its transitive closure is denoted by $\red^+$, and $\red^*$ is
the reflexive-transitive closure.}  
Interesting are the reduction rules
for recursion and corecursion:
\[
\begin{array}[t]{lll@{\hspace{4ex}}l}
  \tfixmu_n\, s\, t_{1..n}\, (\tin\,t) & \red & 
    s\, (\tfixmu_n\, s)\, t_{1..n}\, (\tin\,t) \\
  \tout\,(\tfixnu_n\, s\, t_{1..n}) & \red & 
    \tout\,(s\, (\tfixnu_n\, s)\, t_{1..n})
\end{array}
\]
A recursive function is only unfolded if its recursive argument is a
value, \ie, of the form $\tin\,t$.  This condition is required to
ensure strong normalization; it is present in the work of Mendler
\cite{mendler:lics}, \gim\ \cite{gimenez:strec}, Barthe \etal\
\cite{gimenez:typeBased}, and the author \cite{abel:rairo04}.  Dually,
corecursive functions are only unfolded on demand, \ie, in an
evaluation context, the matching one being $\tout\,\_$.

\longversion{
\begin{figone}{\columnwidth}{\Fhat: Judgements.\label{fig:judge}}
\[
\begin{array}{ll}
p \leq p' 
  & \mbox{polarity ordering} \\
\Delta \der F : \kappa
  & \mbox{kinding} \\
\Delta \der F = F' : \kappa
  & \mbox{constructor equality} \\
\Delta \der F \leq F' : \kappa
  & \mbox{higher-order subtyping} \\
t \red t'
  & \mbox{reduction} \\
\Gamma \der t : A  
  & \mbox{typing} \\
\Gamma \der A \adm{\tfixnabla_n\!}
  & \mbox{admissible recursion type} \\
\end{array}
\]
\end{figone}

Figure~\ref{fig:judge} lists the basic judgements of $\Fhat$, their
rules can be found in the appendix. }%
As pointed out in the
introduction, recursion is introduced by the rule
\begin{gather*}
\nru{\rtyrec
   }{\Gamma \der A \adm{\tfixnabla_n\!} \qquad
     \Gamma \der a : \tord 
   }{\Gamma \der \tfixnabla_n : (\forall \vari\of\tord.\,
     A\,\vari\to A\,(\vari+1)) \to A\,a}
.
\end{gather*}
Herein, $\nabla$ stands for $\mu$ or $\nu$, and the judgement $A
\adm{\tfixnabla_n\!}$ (see Def.~\ref{def:synadm}) 
ensures that type $A$ is admissible for
(co)recursion, as discussed in the introduction.  In this article, we
will find out which types are admissible.

\begin{exa}
  Now we can code the example from the introduction in $\Fhat$, with a
  suitable coding of $\ttrue$, $\tfalse$ and $\wedge$.
\[
\begin{array}{lll}
  \teqGRose & : & (\forall A.\, \Eq\,A \to \Eq\,(F A)) \to \forall A.\,
     \Eq\,A \to \forall \vari.\, \Eq\,(\GRose^\vari F A) \\
  \teqGRose & := & \lambda \veqF \lambda \veqA. \\
  && \fixmu_0 \lambda \veq \lambda t_1 \lambda t_2.\,
  \tcase\,(\tout\,t_1) \\
  && \quad
  \begin{array}[t]{l}
    (\lambda \_.\, \tcase\,(\tout\,t_2)\,
                     (\lambda \_.\,\ttrue)\,
                     (\lambda n_2.\,\tfalse)) \\
    (\lambda n_1.\, \tcase\,(\tout\,t_2)\,(\lambda \_.\,\tfalse) \\
    \quad (\lambda n_2.\,
    \begin{array}[t]{l}
        (\veqA\,(\tfst\,n_1)\,(\tfst\,n_2)) \ \wedge \\
        (\veqF\, \veq\, (\tsnd\,n_1)\, (\tsnd\,n_2))))  
    \end{array}
  \end{array}
\end{array}
\]
  Typing succeeds, by the following assignment of types to variables:
\[
\begin{array}{lll}
  \veqF & : & \forall A.\, \Eq\,A \to \Eq\,(F A) \\
  \veqA & : & \Eq\,A \\
  \veq  & : & \Eq\,(\GRose^\vari F A) \\
\end{array}
\qquad
\begin{array}{lll}
  t_1,t_2 & : & \GRose^{\vari+1} F A \\
  \_      & : & 1 \\
  n_1,n_2 & : & A \times F\,(\GRose^\vari F A) \\
\end{array}
\]
\end{exa}
More examples, including programs over heterogeneous types, can be
found in the author's thesis \cite{abel:PhD}.

\section{Semantics}
\label{sec:sem}

Hughes, Pareto, and Sabry \cite{pareto:sizedtypes} give a
domain-theoretic semantics of sized types.  We, however, follow Barthe
\etal\ \cite{gimenez:typeBased} and interpret types as sets of
terminating \emph{open} expressions and show that any reduction sequence
starting with a well-typed expression converges to a normal form.
This is more than showing termination of programs (closed
expressions); our results can be applied to partial evaluation and
testing term equality in type-theoretic proof assistants.

The material in this section is quite technical, but provides the
necessary basis for our considerations in the following sections.  The
reader may browse it, take a closer look at the interpretation of
types (Sec.~\ref{sec:tyint}) and then continue with
Section~\ref{sec:semicont}, coming back when necessary.

Let $\S$ denote the set of strongly normalizing terms.  We interpret a
type $A$ as a semantic type $\den A \subseteq \S$, and the function
space is defined extensionally:
\[
  \den{A \to B} = \{ r \mid r\,s \in \den B \mbox{ for all } s \in
  \den A \} .
\]
As main theorem, we show that given a well-typed term $x_1\of A_1,
\dots x_n \of A_n \der t : C$ and replacements $s_i \in \den{A_i}$ for
each occurring variable $x_i$, the substitution $\subst{\vec s}{\vec
  x}t$ inhabits $\den C$.  The proof proceeds by induction on the
typing derivation, and in the $\lambda$-case (here simplified)
\[
  \ru{x : A \der t : B
    }{\der \lambda x t : A \to B}
\]
it suffices to show $(\lambda x t)\,s \in \den B$ for any $s \in \den
A$.  However, by induction hypothesis we know only $\subst s x t \in
\den B$.  We therefore require semantic types to be closed under
weak head expansion to make this case go through.

Since we are interested in normalization of open terms, we need to set
aforementioned replacements $s_i$ to variables $x_i$.  This is
possible if each semantic type contains all variables, which has to
be generalized to all neutral terms, i.e. terms $E[x]$ with a variable
in evaluation position.  These observations motivate our definition of
semantic types.

\para{Semantic types}
We define \emph{safe} (weak head)
reduction $\rhd$ by the following axioms.  The idea is that semantic types
are closed under $\rhd$-expansion.
\[
\begin{array}[t]{l@{\hspace{1ex}}l@{\hspace{1ex}}l@{\hspace{1ex}}l}
  (\lambda x t)\, s    & \rhd & \subst s x t
    & \mbox{ if } s \in \S \\
  \tfst\,(\tpair\,r\,s)    & \rhd & r
    & \mbox{ if } s \in \S \\
  \tsnd\,(\tpair\,r\,s)    & \rhd & s
    & \mbox{ if } r \in \S \\
  \tout\,(\tin\,r) & \rhd & r \\
\end{array}
\qquad
\begin{array}[t]{l@{\hspace{1ex}}l@{\hspace{1ex}}l@{\hspace{1ex}}l}
  \tcaseraw\,(\tinl\,r) & \rhd & \lambda x\lambda y.\,x\,r  & (*)\\
  \tcaseraw\,(\tinr\,r) & \rhd & \lambda x\lambda y.\,y\,r  & (*)\\
  \tfixmu_n s\ t_{1..n}\, (\tin\,r) & \rhd & 
    s\, (\tfixmu_n s)\, t_{1..n}\, (\tin\,r)\\
  \tout\,(\tfixnu_n s\, t_{1..n}) & \rhd & 
    \tout\, (s\, (\tfixnu_n s)\, t_{1..n}) \\
\end{array}
\] 
Side condition (*):  $x,y \not\in\FV(r)$.
Additionally, we close safe reduction under evaluation contexts and
transitivity:
\[
\begin{array}{l@{\hspace{1ex}}l@{\hspace{1ex}}l@{\hspace{4ex}}l}
  E(t) & \rhd & E(t') 
    & \mbox{ if } t \rhd t' \\
  t_1  & \rhd & t_3 
    & \mbox{ if } t_1 \rhd t_2 \mbox{ and } t_2 \rhd t_3 \\  
\end{array}
\]
One-step safe reduction is deterministic, hence, if $r \rhd s$ and $r
\rhd t$ then either $s=t$ or $s \rhd t$ or $t \rhd s$.  
\[
  \V := \{ v, E(x) \mid v \mbox{ value}, E \mbox{ evaluation
  context} \}
\]
is the set of $\rhd$-normal forms, not counting junk terms like
$\tfst\,(\lambda x t)$.

The relation is defined such that $\S$ is closed under
$\rhd$-expansion, meaning $t \rhd t' \in \S$ implies $t \in \S$.
In other words, $\rhd$ used in the expansion direction does not
introduce diverging terms. 
Let $\triA$ denote the closure of term set $\A$ under
$\rhd$-expansion.   In general, the \emph{closure} of term set $\A$ is
defined as
\[
  \clos\A = \tri(\A \cup \{ E(x) \mid x \mbox{ variable}, E(x) \in \S\}).
\]
Closure preserves strong normalization:
  If $\A \subseteq \S$ then $\clos\A \subseteq \S$.
A term set is \emph{closed} if $\clos\A = \A$.
The least closed set is the set of neutral terms $\N := \clos\emptyset
\not= \emptyset$.
Intuitively, a neutral term never reduces to a value, it necessarily
has a free variable, and it can be
substituted into any term without creating a new redex.  
A term set $\A$ is \emph{saturated} if $\A$ is closed and $\N
\subseteq \A \subseteq \S$ (this makes sure that $\A$ contains all
variables).  A saturated set is called a \emph{semantic type}. 

\para{Interpretation of kinds}
When types are interpreted as sets of terms, the easiest
interpretation of type constructors are set-theoretical operators on
term sets, or as we go higher-order, on operators. 

The saturated sets form a complete lattice
$\den\ttype$ with least element $\lbot[\ttype] := \N$ and greatest
element $\ltop[\ttype] := \S$. It is ordered by inclusion ${\lsub[\ttype]}
:= {\subseteq}$ and has set-theoretic infimum $\linf[\ttype] := \bigcap$
and supremum $\lsup[\ttype] := \bigcup$.  Let $\den\tord := \O$ where
$\O = [0;\ltop[\tord]]$ is an initial segment of the set-theoretic
ordinals.  With the usual ordering on
ordinals, $\O$ constitutes a complete lattice as well.
For lattices $\LL$ and $\LL'$, let $\LL \toco \LL'$ denote the space of
monotone functions from $\LL$ to $\LL'$ and $\LL \tocontra \LL'$ the
space of antitone ones.
The mixed-variant function kind $\den{\pin\kappa\to\kappa'}$ is
in\-ter\-pre\-ted as set-theoretic function space 
$\den{\kappa} \to \den{\kappa'}$; the covariant
function kind $\pco\kappa \to \kappa'$ denotes
the monotonic function space $\den\kappa \toco \den{\kappa'}$ 
and the contravariant kind $\pcontra \kappa
\to \kappa'$ the antitonic space $\den\kappa \tocontra \den{\kappa'}$.
For all function kinds, ordering is defined pointwise:
$\F \lsub[p\kappa\to\kappa'] \F' \defiff \F(\G) \lsub[\kappa'] \F'(\G)$
for all $\G \in \den\kappa$.  Similarly, $\lbot[p\kappa\to\kappa'](\G)
:= \lbot[\kappa']$ is defined pointwise, and so are
$\ltop[p\kappa\to\kappa']$, $\linf[p\kappa\to\kappa']$,  and $\lsup[p\kappa\to\kappa']$.

\para{Limits and iteration}
Inductive types $\den{\mu^a F}$ are constructed by iterating
the operator $\den F$ $\den a$-times, starting with the least
semantic type $\lbot$.  At limit ordinals, we take the supremum.  If
$\den a$ is big enough, latest if $\den a = \ltop[\tord]$, the least
fixed-point is reached, but our type system also provides notation for the
approximation stages below the fixed-point.  For coinductive types, we
start with the biggest semantic type $\ltop$ and take the infimum at
limits.  It is possible to unify these two forms of iteration, by
taking the $\limsup$ instead of infimum or supremum at the limits.
The notion of $\limsup$ and iteration can be defined for arbitrary lattices:

In the following $\lambda \in \O$ will denote a limit ordinal. (We
will only consider proper limits, \ie, $\lambda \not= 0$.)  For $\LL$
a complete lattice and $f \in \O \to \LL$ we define:
\[
\begin{array}{lll@{\hspace{4ex}}l}
  \liminf_{\alpha\to\lambda} f(\alpha) & := & \sup_{\alpha_0 < \lambda} \inf_{\alpha_0
    \leq \alpha < \lambda}  f(\alpha) 
\\
  \limsup_{\alpha\to\lambda} f(\alpha) & := & \inf_{\alpha_0 < \lambda} \sup_{\alpha_0
    \leq \alpha < \lambda}  f(\alpha) 
\\
\end{array}
\]
Using $\inf_\lambda f$ as shorthand for $\inf_{\alpha<\lambda}
f(\alpha)$, and analogous shorthands for $\sup$, $\liminf$, and $\limsup$,
we have 
$\inf_\lambda f \lsub \liminf_\lambda f \lsub \limsup_\lambda f \lsub
\sup_\lambda f$.  If $f$ is monotone, then even 
$\liminf_\lambda f = \sup_\lambda f$, and if
$f$ is antitone, then $\inf_\lambda f = \limsup_\lambda f$.

If $f \in \LL \to \LL$ and $g \in \LL$, we define transfinite
iteration $f^\alpha(g)$ by recursion on $\alpha$ as follows:
\[
\begin{array}{l@{\hspace{0ex}}lll}
  f^0 & (g) & := & g \\
  f^{\alpha+1} & (g) & := & f(f^\alpha(g)) \\
  f^{\lambda} & (g) & := & \limsup_{\alpha\to\lambda} f^\alpha(g) \\
\end{array}
\]
This definition of iteration works for any $f$, not just monotone ones.
For monotone $f$, we obtain the usual approximants of least and
greatest fixed-points as $\MU^\alpha f = f^\alpha(\bot)$ and
$\NU^\alpha f = f^\alpha(\top)$:  It is easy to check that
$\MU^\lambda f = \sup_{\alpha < \lambda} \MU^\alpha f$ and
$\NU^\lambda f = \inf_{\alpha < \lambda} \NU^\alpha f$, so our
definition coincides with the usual one.

\para{Closure ordinal}
\label{sec:closord}
We can calculate an upper bound for the ordinal $\ltop[\tord]$ at which
all fixed-points are reached as follows:
Let $\uncount_n$ be a sequence of cardinals defined by $\uncount_0 =
|\NN|$ and $\uncount_{n+1} = |\P(\uncount_n)|$.  For a pure kind
$\kappa$, let $|\kappa|$ be the number of $\ttype$s in $\kappa$.
Since $\den\ttype$ consists of countable sets, $|\den\ttype| \leq
|\P(\NN)| = \uncount_1$, and by induction on $\kappa$, 
$|\den\kappa| \leq \uncount_{|\kappa|+1}$.  Since an (ascending or
descending) chain in $\den\kappa$ is shorter than $|\den{\kappa}|$, each
fixed point is reached latest at the $|\den{\kappa}|$th iteration.
Hence, the closure ordinal for all (co)inductive types can be
approximated from above by $\ltop[\tord] = \uncount_\omega$.

This calculation does not work if we allow fixed-points of
constructors involving $\tord$.  Then the closure ordinal of such a
fixed-point would depend on which ordinals are in the semantics of
$\tord$, which in turn would depend on what the closure ordinal for
all fixed-points was---a vicious cycle.  However, I do not see a
practical example where one want to construct the fixed point of a
sized-type transformer $F : (\tord \toin \kappa) \toco (\tord \toin
\kappa)$.  Note that this does not exclude fixed-points inside
fixed-points, such as
\[\BTree^{\vari,\varii} A = \mu^\vari \lambda X.\ 1 + X \times (\mu^\varii \lambda
Y.\, 1 + A \times X \times Y),\] ``B-trees'' of height $< \vari$ with each
node containing $< \varii$ keys of type $A$.

\begin{exa}[Number classes]
Here we show that higher-kinded strictly-positive inductive types may require
strictly higher closure ordinals than strictly-positive inductive types of kind
$\ttype$.  Following Hancock \cite{hancock:numberclasses}, we can
define the number classes as inductive types as follows:
\[
\begin{array}{lllll}%
  \NC_0 & := & \mu^\infty \lambda X.\, 1 
    & \cong & 1 \\
  \NC_1 & := & \mu^\infty \lambda X.\, 1 + (\NC_0 \to X)
    & \cong & \Nat^\infty \\  
  \NC_2 & := & \mu^\infty \lambda X.\, 1 + (\NC_0 \to X) + (\NC_1 \to X)
    & \cong &  \mu^\infty \lambda X.\, 1 + X + (\Nat^\infty \to X)
    \\
  \NC_3 & := 
  & \multicolumn 3 {l} {
      \mu^\infty \lambda X.\, 1 + (\NC_0 \to X) + (\NC_1 \to X) + (\NC_2 \to X)
    } \\
  \multicolumn 1 c \vdots
\end{array}
\]
The second number class $\NC_2$ is also known as \emph{Brouwer ordinals}.
The law behind this scheme is: $\NC_n = \mu^\infty F_n$, where $F_0\,X = 1$ 
and $F_{n+1}\,X = F_n\,X + (\mu^\infty F_n \to X)$. 
Each number class requires a higher closure ordinal, and their limit
is the closure ordinal of all strictly-positive inductive types of kind $\ttype$.
Now let
\[
  \begin{array}{lll}
    \NumClTree & :  & \tord \toco (\ttype \toco \ttype) \toin \ttype \\
    \NumClTree & := & \lambda \vari.\, \mu^\vari \lambda Y \lambda
    F.\, 1 + (\mu^\infty F \to Y\,(\lambda X.\, F\,X + (\mu^\infty F
    \to X))) . \\
  \end{array}
\]  
Then $\NumClTree^\infty\,(\lambda X.\,1)$ is the type of trees
branching over the $n$th number class at the $n$th level.
This example suggests that the closure ordinal of certain strictly positive
inductive types of kind $(\ttype \toco \ttype) \toin \ttype$ is above
the one of the strictly-positive inductive types of kind~$\ttype$.
However, the situation is unclear for non-strictly positive inductive types.
\end{exa}

\para{Interpretation of types} \label{sec:tyint}
For $r$ a term, $e$ an evaluation frame, and $\A$ a term set, let $r
\sapp \A = \{ r\,s \mid s \in \A \}$ and $e^{-1}\A = \{ r \mid e(r) \in
    \A \}$.  
\longversion{
If $e$ is strongly normalizing and $\A$ saturated, then
    $e^{-1}\A$ is again saturated.
}
For saturated sets $\A,\B \in \den\ttype$ we define the following
    saturated sets:
\[
\def\arraystretch{1.5}
\begin{array}[t]{lll@{\hspace{4ex}}l}
  \A \bplus \B & := & \clos{\tinl\sapp\A} \cup \clos{\tinr\sapp\B} \\
  \A \btimes \B & := & (\tfst\,\_)^{-1}\!\A \cap (\tsnd\,\_)^{-1}\!\B \\
  \A \bto \B & := & \bigcap_{s\in\A}\, (\_\,s)^{-1}\!\B \\
\end{array}
\qquad
\begin{array}[t]{lll@{\hspace{4ex}}l}
  \bone & := & \clos{\{\ttuple{}\}} \\
  \A^\mu & := & \clos{\tin\sapp\A} \\
  \A^\nu & := & (\tout\,\_)^{-1}\A
\end{array}
\]
The last two notations are lifted pointwise to operators $\F \in
\den{p\kappa\to\kappa'}$ by setting $\F^\nabla(\G) = (\F(\G))^\nabla$,
where $\nabla \in \{\mu,\nu\}$.
\begin{rem}
  Our definition of product and function space (inspired by
  Vouillon \cite{vouillon:subtypingUnion}) makes it
  immediate that $\btimes$ and $\bto$ operate on saturated sets.
  But it is just a
  reformulation of the usual $\A \btimes \B = \{ r \mid \tfst\,r \in
  \A \mbox{ and } \tsnd\,r \in \B \}$ and
  $\A \to \B = \{ r \mid r\,s \in \B \mbox{ for all } s \in \A \}$.

  Notice that the \emph{finitary} or (in the logical sense) 
  \emph{positive} connectives $1$, $+$, and $\mu$ are
  defined via introductions, while the \emph{infinitary} or
  \emph{negative} 
  connectives $\to$ and $\nu$ are defined via eliminations.  (The
  binary product $\times$ fits in either category.)
\end{rem}
For a constructor constant $C\of\kappa$, the semantics $\den C \in
\den\kappa$ is defined as follows:
\[
\def\arraystretch{1.3}
\begin{array}[t]{lll}
\den{+}(\A,\B \in \den\ttype) & := & \A \bplus \B \\
\den{\times}(\A,\B \in \den\ttype) & := & \A \btimes \B \\
\den{\to}(\A,\B \in \den\ttype) & := & \A \bto \B \\
\den{\mu_\kappa}(\alpha)(\F \in \den\kappa \toco\den\kappa) & := &
\MU^\alpha \F^\mu \\
\den{\nu_\kappa}(\alpha)(\F \in \den\kappa \toco\den\kappa) & := & \NU^\alpha \F^\nu \\
\den{\forall_\kappa}(\F \in \den\kappa \to \den\ttype) & := &
\bigcap_{\G \in [\![\kappa]\!]} \F(\G) \\
\end{array}
\qquad
\begin{array}[t]{lll}
\den{1} & := & \bone \\
\den{\infty} & := & \ltop[\tord] \\
\den{\s}(\ltop[\tord]) & := & \ltop[\tord] \\
\den{\s}(\alpha < \ltop[\tord]) & := & \alpha + 1 \\
\end{array}
\]
\shortversion{
We extend this semantics to constructors $F$ in the usual way, such
that if $\Delta \der F : \kappa$ and $\theta(X) \in \den{\kappa'}$ for
all $(X\of p\kappa') \in \Delta$, then $\Den F \theta \in \den\kappa$.
}
\longversion{
This semantics is extended to arbitrary constructors in the usual way.
Let $\U = \bigcup_\kappa \den{\kappa}$.
For a valuation $\theta$ which partially 
maps constructor variables $X$ to their
interpretation $\G \in \U$, we define the partial map
$\Den{-}\theta$ from constructors $F$ to their interpretation in $\U$
by recursion on $F$.
\[
\begin{array}{lll}
\Den C \theta & := & \den C \\
\Den X \theta & := & \theta (X) \\
\Den{F\,G}\theta & := & \Den{F}\theta(\Den{G}\theta) \\[2ex]
\Den{\lambda X\ofkinddot{\kappa} F}\theta & := & 
  \Bigl\{
    \begin{array}{l@{\hspace{2ex}}l}
       \F & \mbox{if } \F \in \den{\kappa}\to\den{\kappa'} \mbox{
         for some } \kappa' \\
       \mbox{undef.} & \mbox{else}
    \end{array} \\
& & \mbox{where } \F(\G \in \den{\kappa}) := \Den{F}{\update \theta X \G}
\end{array}
\]
In the last clause, $\F$ is a partial function from $\den\kappa$ to
$\U$.

The interpretation $\Den{F}\theta$ is well-defined for well-kinded
$F$, and these are the only constructors we are interested in, but we
chose to give a (possibly undefined) meaning to all constructors.  If
one restricts the interpretation to well-kinded constructors, one has
to define it by recursion on kinding derivation and show coherence:
If a constructor has two kinding derivations ending in the same kind,
then the two interpretations coincide.  This alternative requires
a bit more work than our choice.

\begin{lem}[Basic properties of interpretation] \hfill
  \begin{enumerate}
  \item Relevance: If $\theta(X) = \theta'(X)$ for all $X \in \FV(F)$, then $\Den
    F \theta = \Den F {\theta'}$.
  \item Substitution: $\Den{\subst G X F}\theta = \Den F {\update
      \theta X {\Den G \theta}}$.
  \end{enumerate}
\end{lem}
\proof
Each by induction on $F$.  For (2), consider
case $F = \lambda Y \ofkinddot\kappa F'$.  \Wlog, $Y \not\in\FV(G)$. 
By induction hypothesis,
\[\F(\H) := \Den {\subst G X F} {\update \theta Y \H} 
          = \Den F {\update
  {\update \theta Y \H} X {\Den G {\update \theta Y \H}}} = \Den F
{\update {\update \theta X {\Den G \theta}} Y \H},\] 
using (1) on $G$.  Hence, $\Den {\subst G X {(\lambda Y \ofkinddot
    \kappa F)}} \theta = \Den {\lambda Y \ofkinddot \kappa F}{\update
  \theta X {\Den G \theta}}$.
\qed
Although the substitution property holds even for ill-kinded
constructors, we only have for well-kinded constructors
that $\Den {(\lambda X \ofkinddot \kappa F)\,G}\theta = \Den{\subst G
  X F}\theta$.  In general, the left hand side is less defined than the right hand
side, \eg, $\Den{(\lambda X \ofkinddot \ttype 1)\,\infty}\theta$ is
undefined, whereas the interpretation $\Den 1 \theta$ of its
$\beta$-reduct is well-defined.  In the following we show
that for well-kinded constructors the interpretation is well-defined
and invariant under $\beta$.
\begin{thm}[Soundness of kinding, equality, and subtyping for constructors]
   Let $\theta \lsub \theta' \in \den \Delta$, meaning that for all
   $(X \of p\kappa') \in \Delta$ it holds that
   $\G := \theta(X) \in \den{\kappa'}$ and
    $\G' := \theta'(X) \in \den{\kappa'}$, and $\G = \G'$ if $p =
    \pin$, $\G \lsub \G'$ if $p = \pco$, and $\G' \lsub \G$ if $p =
    \pcontra$.
    \begin{enumerate}
    \item If $\Delta \der F : \kappa$ then $\Den F \theta \lsub \Den F
      {\theta'} \in \den\kappa$.
    \item If $\Delta \der F = F' : \kappa$ then $\Den F \theta \lsub \Den
    {F'}{\theta'} \in \den\kappa$.
    \item If $\Delta \der F \leq F' : \kappa$ then $\Den F \theta
      \lsub \Den {F'}{\theta'} \in \den\kappa$. 
    \end{enumerate}
\end{thm}
\proof
Simultaneously by induction on the derivation.
\qed
}

Now we can compute the semantics of types, \eg, $\Den{\Nat^\vari}{(\vari
  \mapsto \alpha)} = \NAT^\alpha = \MU^\alpha
(\X \mapsto (\bone \bplus \X)^\mu)$.  Si\-mi\-lar\-ly, the semantic
versions of $\List$, $\Stream$, etc.\ are denoted by $\LIST$,
$\STREAM$, etc.

\para{Semantic admissibility and strong normalization}  For the
main theorem to follow, we assume semantic soundness of our yet
to be defined syntactical criterion of admissibility (Def.~\ref{def:synadm}).
\begin{asm}[Semantic admissibility]\label{asm:semadm}
If $\Gamma \der A \adm{\tfixnabla_n\!}$ and $\theta(X) \in \den{\kappa}$
for all $(X\of \kappa) \in \den\Gamma$ then $\A := \Den A \theta \in
\den\tord \to \den\ttype$ has the
following properties:
\begin{enumerate}
\item\label{it:shape}
 Shape: $\A(\alpha) = \bigcap_{k\in K} \B_1(k,\alpha) \bto \dots
\bto \B_n(k,\alpha) \bto \B(k,\alpha)$ for some $K$ and some $\B_1, \dots, \B_n,
\B \in K \times \den\tord \to \den\ttype$.  In case $\nabla=\mu$, $\B(k,\alpha) = \calI(k,\alpha)^\mu \bto
\C(k,\alpha)$ for some $\calI, \C$.  Otherwise, $\B(k,\alpha) =
\C(k,\alpha)^\nu$ for some $\C$.
\item\label{it:bottom}
 Bottom-check: $\calI(k,0)^\mu = \lbot[\ttype]$ in case $\nabla=\mu$
  and $\C(k,0)^\nu = \ltop[\ttype]$ in case $\nabla=\nu$.
\item\label{it:limit}
 Limit-check: $\inf_{\alpha < \lambda} \A(\alpha) \subseteq \A(\lambda)$
  for all limit ordinals $\lambda \in \den\tord\setminus\{0\}$.
\end{enumerate}  
\end{asm}
In case of recursion $(\nabla = \mu)$, the condition (\ref{it:shape})
ensures that $\tfixmu_n s$ really produces a function whose $n+1$st
argument is of something that looks like an inductive type
($\calI(k,\alpha)^\mu$).  The function can be polymorphic, therefore
the intersection $\bigcap_{k\in K}$ over an index set $K$.
Condition (\ref{it:bottom}) requires $\calI$
to exhibit at least for $\alpha = 0$ the behavior of an inductive
type: $\calI(k,0)^\mu = \lbot[\ttype]$, which is equivalent to
$\calI(k,0) = \emptyset$.  The technical condition (\ref{it:limit}) is
used in the following theorem and will occupy our attention for the
remainder of this article.  In case of corecursion $(\nabla = \nu)$,
condition (\ref{it:shape}) ensures that $\tfixnu_n s$ maps $n$
arguments into something like a coinductive type ($\C(k,\alpha)^\nu$),
which needs to cover
the whole universe $\ltop[\ttype]$ of terms for $\alpha = 0$.

Now we show soundness of our typing rules, which entails strong normalization.
Let $t\theta$ denote the simultaneous substitution of $\theta(x)$ for
each $x \in \FV(t)$ in $t$.
\begin{thm}[Soundness of typing]
  Assume that the judgement $\Gamma \der A \adm{\tfixnabla_n\!}$ is
  sound, as stated above.
  Let $\theta(X) \in \den\kappa$ for all $(X\of\kappa) \in \Gamma$ and
  $\theta(x) \in \Den A \theta$ for all $(x\of A)\in \Gamma$.
  If\/ $\Gamma \der t : B$ then $t\theta \in \Den B \theta$.
\end{thm}
\longversion{
\begin{proof}
  By induction on the typing derivation.  We consider the recursion
  rule ($\rtyrec$ for $\nabla=\mu$).  
\[
\nru{\rtyrec
   }{\Gamma \der A \adm{\tfixmu_n\!} \qquad
     \Gamma \der a : \tord 
   }{\Gamma \der \tfixmu_n : (\forall \vari\of\tord.\,
     A\,\vari\to A\,(\vari+1)) \to A\,a}
\]
  By hypothesis, $\A := \Den A
  \theta \in \den\tord \to \den\ttype$ is admissible, and $\alpha :=
  \Den a \theta \in \den\tord$.  Assume an $s \in \Den{\forall
    \vari\of\tord.\ A\,\vari \to A\,(\vari + 1)}\theta \subseteq
  \bigcap_{\beta<\top^\tord} \A(\beta) \bto \A(\beta+1)$.  We show
  $\tfixmu_n\,s \in \A(\alpha)$ by transfinite induction on $\alpha$.

  In the base case $\alpha=0$, by Assumption~\ref{asm:semadm} we have
  $\A(0) = \bigcap_{k\in K} \B_{1..n}(k,0) \bto
  \lbot[\ttype] \bto \C(k,0)$.  We assume $k\in K$, $t_i \in \B_i(k,0)$, then
  $e(r) := \tfixmu_n\,s\,t_{1..n}\,r$ is a strongly normalizing
  evaluation frame.  Since each $r \in \lbot[\ttype] = \N$ is neutral,
  we have $e(r) \in \N \subseteq \C(k,0)$.

  In the step case, $\A(\alpha+1) = \bigcap_{k\in K} \B_{1..n}(k,\alpha+1) \bto
  \calI(k,\alpha+1)^\mu \bto \C(k,\alpha+1)$.  We assume $k\in K$,
  $t_i \in \B_i(k,\alpha+1)$, and $r \in \calI(k,\alpha+1)^\mu$, which
  means that either $r$ is neutral---then we continue as in the
  previous case---or $r \rhd \tin\,r'$.  Now $\tfixmu_n\,s\,\vec t\,r \rhd
  \tfixmu_n\,s\,\vec t\,(\tin\,r') \rhd s\,(\tfixmu_n\,s)\,\vec
  t\,(\tin\,r')$.  The last term inhabits $\C(k,\alpha+1)$, since $\tfixmu_n\,s \in
  \A(\alpha)$ by induction hypothesis and therefore $s\,(\tfixmu_n\,s)
  \in \A(\alpha+1)$.

  Finally, in the limit case, $\tfixmu_n\,s \in \A(\alpha)$ for all
  $\alpha<\lambda$ by induction hypothesis.  Since
  $\bigcap_{\alpha<\lambda} \A(\alpha) = \inf_{\alpha<\lambda}
  \A(\alpha) \subseteq
  \A(\lambda)$ by Assumption~\ref{asm:semadm}, we are done.
\end{proof}
}
\begin{cor}[Strong normalization]
  If $\Gamma \der t : B$ then $t$ is strongly normalizing.
\end{cor}
\longversion{
\begin{proof}
  From soundness of typing, taking
  a valuation $\theta$ with $\theta(x) = x$ for all term variables $x$
  and $\theta(X) = \ltop[\kappa]$ for all $(X \of p\kappa) \in \Gamma$.
\end{proof}
}

\section{Semi-Continuity}
\label{sec:semicont}

As motivated in the introduction, only types $\C \in \den{\tord} \to
\den{\ttype}$ which satisfy the limit-check 
$\inf_\lambda \C \lsub \C(\lambda)$ can be
admissible for recursion.  In this section, we develop a compositional
criterion for admissible types.  The limit-check
itself is not compositional since it does not sensibly distribute over
function spaces:  To show \(\inf_{\alpha < \lambda} (\A(\alpha) \bto
  \B(\alpha)) \lsub \A(\lambda) \bto \B(\lambda)\) from $\inf_\lambda
  \B \lsub \B(\lambda)$ requires
$\A(\lambda) \lsub \inf_\lambda \A$, which is not even true for $\A(\alpha)
  = \NAT^\alpha$
at limit $\omega$.  However, the criterion $\limsup_\lambda \C \lsub
\C(\lambda)$ entails the limit-check, and it distributes reasonably
over the function space:
\begin{prop}\label{prop:fun}
  If $\A(\lambda) \lsub \liminf_\lambda \A$ and $\limsup_\lambda \B
  \lsub \B(\lambda)$ then $\limsup_\lambda (\A(\alpha) \bto
  \B(\alpha)) \lsub \A(\lambda) \bto \B(\lambda)$.
\end{prop}
\noindent
This proposition will reappear (and be proven) as Cor.~\ref{cor:fun}.
Note that $\NAT^\omega = \liminf_\omega \NAT$, hence, $\A(\alpha) =
\NAT^\alpha$ can now serve as the domain of an admissible function
space, which is the least we expect.

The conditions on $\A$ and $\B$ in the lemma are established
mathematical terms: They are subconcepts of continuity.  In this
article, we consider only functions $f \in \O \to \LL$ from ordinals
into some lattice $\LL$.  For such $f$, the question whether $f$ is
continuous in point $\alpha$ only makes sense if $\alpha$ is a limit
ordinal, because only then there are infinite non-stationary sequences
which converge to $\alpha$; and since every strictly decreasing
sequence is finite on ordinals (well-foundedness!), it only makes
sense to look at \emph{ascending} sequences, \ie, approaching the
limit from the left.  Hence, function $f$ is \emph{upper
  semi-continuous} in $\lambda$, if $\limsup_\lambda f \lsub
f(\lambda)$, and \emph{lower semi-continuous}, if $f(\lambda) \lsub
\liminf_\lambda f$.  If $f$ is both upper and lower semi-continuous
in $\lambda$, then it is continuous in $\lambda$ (then upper and lower
limit coincide with $f(\lambda)$).

In the following we identify sufficient criteria for sum, product,
function, inductive, and coinductive types to be semi-continuous.

\para{Semi-continuity from monotonicity} \bla

Obviously, any monotone function is upper semi-con\-tinu\-ous, and any
antitone function is lower semi-continuous.  Now consider a monotone
$f$ with $f(\lambda) = \sup_\lambda f$, as it is the case for an
inductive type $f(\alpha) = \MU^\alpha \F$ (where $\F$ does not depend
on $\alpha$).  Since for monotone $f$, $\sup_\lambda f =
\liminf_\lambda f$, $f$ is lower semi-continuous.  This criterion can
be used with Prop.~\ref{prop:fun} to show upper semi-continuity of
function types with inductive domain, such as
$\Eq(\GRose^\vari\,F A)$ (see introduction) and, \eg,
\[
  \C(\alpha) = \NAT^\alpha \bto \LIST^\alpha(\A) \bto \C'(\alpha)
\]
where $\C'(\alpha)$ is any monotonic type-valued function, for
instance, $\LIST^\alpha(\NAT^\alpha)$, and $\A$ is some constant type:
The domain types, $\NAT^\alpha$ and $\LIST^\alpha(\A)$, are lower
semi-continuous according the just established criterion and the
monotonic codomain $\C'(\alpha)$ is upper semi-continuous, hence, 
Prop.~\ref{prop:fun} proves upper semi-con\-ti\-nu\-i\-ty of $\C$.  
Note that this criterion fails us if
we replace the domain $\LIST^\alpha(\A)$ by
$\LIST^\alpha(\NAT^\alpha)$, or even $\MU^\alpha (\F(\NAT^\alpha))$ for
some monotone $\F$, since it is not immediately obvious that
\begin{displaymath}
    \MU^\omega(\F(\NAT^\omega))
 =  
    \sup_{\alpha<\omega} \MU^\alpha (\F(\sup_{\beta<\omega} \NAT^\beta))  
 \stackrel{?}{=} 
    \sup_{\gamma<\omega} \MU^\gamma (\F(\NAT^\gamma)).
\end{displaymath}
However, domain types where one indexed inductive type is inside
another inductive type are useful in practice, see
Example~\ref{ex:bf}.  Before we consider
lower semi-continuity of such types, let us consider the dual case.

For $f(\alpha) = \NU^\alpha \F$, $\F$ not dependent on $\alpha$, $f$
is antitone and $f(\lambda) := \limsup_\lambda f = \inf_\lambda f$,
hence, $f$ is continuous in all limits.
This establishes upper semi-continuity of a type
involved in stream-zipping,
\[
  \STREAM^\alpha(\A) \bto \STREAM^\alpha(\B) \bto \STREAM^\alpha(\C) .
\]
However, types like $\STREAM^\alpha(\NAT^\alpha)$ are not yet covered, but now
we will develop concepts that allow us to look inside (co)inductive types.

\para{Simple semi-continuous types}
First we will investigate how disjoint sum, product, and function
space operate on semi-continuous types.

\begin{defi}
Let $f
\in \LL \to \LL'$.  We say \emph{$\limsup$ pushes through $f$}, or $f$
is \emph{$\limsup$-pushable}, if for all $g \in \O \to \LL$,
\[
  \limsup_{\alpha\to\lambda} f(g(\alpha)) \lsub f(\limsup_\lambda g).
\]
Analogously, $f$ is \emph{$\liminf$-pullable}, or \emph{$\liminf$ can
  be pulled out of $f$}, if for all $g$,
\[
  f(\liminf_\lambda g) \lsub \liminf_{\alpha\to\lambda} f(g(\alpha)).
\]
These notions extend straightforwardly to $f$s with several arguments.  
\end{defi}

\begin{lem}[Facts about limits]\label{lem:fact} \bla
\begin{enumerate}
\item\label{it:pullsup}
 $\textstyle     \sup_{i\in I} \liminf_{\alpha \to \lambda} h(\alpha,i) \lsub
    \liminf_{\alpha \to \lambda} \sup_{i\in I} h(\alpha,i)$.
\item\label{it:pushinf}
 $\textstyle    \limsup_{\alpha \to \lambda} \inf_{i\in I} h(\alpha,i) \lsub
    \inf_{i\in I} \limsup_{\alpha \to \lambda} h(\alpha,i)$.
\item\label{it:pushinfdep}
 $\textstyle    \limsup_{\alpha \to \lambda} \inf_{i\in I(\alpha)} h(\alpha,i) \lsub
    \inf_{i\in \liminf_\lambda I} \limsup_{\alpha \to \lambda} h(\alpha,i)$.
\qed
\end{enumerate}
\end{lem}
\noindent
Fact~(\ref{it:pushinf}) states that $\limsup$ pushes through infimum, 
setting $\LL = K \to \LL'$ for some set $K \supseteq I$, 
$f(g') = \inf_{i \in I} g'(i)$, and $g(\alpha)(i) =
h(\alpha,i)$ in the above definition. 
Thus, universal quantification is $\limsup$-pushable, which
justifies rule $\rcontforall$ in Figure~\ref{fig:semi} (see
Sect.~\ref{sec:der}).  The dual fact~(\ref{it:pullsup})
expresses that $\liminf$ can be pulled out of a supremum.

Fact~(\ref{it:pushinfdep}) is a generalization of (\ref{it:pushinf}) which
we will need to show semi-continuity properties of the function space.
\proof
In the following proof of (\ref{it:pushinfdep}), let all ordinals
range below $\lambda$.  First we derive
\[
\begin{array}{rcl@{\quad}l}
  h(\alpha,i) & \lsub & \sup_{\alpha \geq \alpha_0} h(\alpha,i) 
    & \mbox{for } \alpha \geq \alpha_0 \\
  \inf_{i\in I(\alpha)} h(\alpha,i) & \lsub & \inf_{i \in \bigcap_{\alpha
      \geq \alpha_0} I(\alpha)} \sup_{\alpha \geq \alpha_0}
  h(\alpha,i) 
    & \mbox{for } \alpha \geq \alpha_0 \\
  \sup_{\alpha \geq \alpha_0} \inf_{i\in I(\alpha)} h(\alpha,i) 
    & \lsub & \inf_{i \in \inf_{\alpha
      \geq \alpha_0} I(\alpha)} \sup_{\alpha \geq \alpha_0}
  h(\alpha,i) . \\
\end{array}
\]
($I(\alpha)$ is a set, so intersection = infimum.)
Secondly, note that 
\[ \textstyle
\inf_{\alpha_0} \inf_{i \in J(\alpha_0)}
g(\alpha_0,i) = \inf_{\alpha_0, i \in J(\alpha_0)} \inf_{\alpha_0}
g(\alpha_0,i) = \inf_{i \in \sup_{\alpha_0} J(\alpha_0)}
\inf_{\alpha_0} g(\alpha_0,i).
\]  
With $g(\alpha_0,i) := \sup_{\alpha
  \geq \alpha_0} h(\alpha,i)$ and $J(\alpha_0) := \inf_{\alpha \geq
  \alpha_0} I(\alpha)$ we finally derive
\[
\begin{array}{rcl}
 \limsup_{\alpha \to \lambda} \inf_{i\in I(\alpha)} h(\alpha,i) 
    & = & \\
 \inf_{\alpha_0} \sup_{\alpha \geq \alpha_0} \inf_{i\in I(\alpha)} h(\alpha,i) 
    & \lsub &
 \inf_{\alpha_0} \inf_{i \in \inf_{\alpha
      \geq \alpha_0} I(\alpha)} \sup_{\alpha \geq \alpha_0}
  h(\alpha,i) \\
   & = & \inf_{i \in \sup_{\alpha_0} \inf_{\alpha \geq \alpha_0} I(\alpha)}
\inf_{\alpha_0} \sup_{\alpha \geq \alpha_0} h(\alpha,i) \\
   & = & \inf_{i \in \liminf_{\lambda} I}
\limsup_{\alpha \to \lambda} h(\alpha,i) . \\
\end{array}
\]
\qed
\begin{lem}[$\liminf$ can be pulled out of the building blocks of saturated sets]
  \hfill \\
$
  \begin{array}{@{\ \ \quad}l@{~}l@{~}l@{~}l}
  (1) & r \sapp \liminf_\lambda \A & \subseteq &
      \liminf_{\alpha \to \lambda} (r \sapp \A(\alpha)). \\
  (2) & e^{-1}(\liminf_\lambda \A) & \subseteq & 
      \liminf_{\alpha \to \lambda} e^{-1}(\A(\alpha)). \\
  (3) & \liminf_\lambda \A \cap \liminf_\lambda \B & \subseteq &
      \liminf_{\alpha \to \lambda} (\A(\alpha) \cap \B(\alpha)). \\
  (4) & \tri(\liminf_\lambda \A)  & \subseteq &
      \liminf_{\alpha \to \lambda} \tri(\A(\alpha)).
  \end{array}
$
\end{lem}
\proof
All propositions have easy proofs.  Let all introduced ordinals range below
$\lambda$.  For proposition~(3), assume $r \in
\liminf_\lambda \A \cap \liminf_\lambda \B$, which means that there
are $\alpha_0,\beta_0$
such that $r \in \A(\alpha)$ for all
$\alpha$ with $\alpha_0 \leq \alpha$
and $r \in B(\beta)$
for all $\beta$ with $\beta_0 \leq \beta$%
.  We have to show
that there exists $\gamma_0$
such that $r \in \A(\gamma)
\cap \B(\gamma)$ for all $\gamma$ with $\gamma_0 \leq \gamma$%
.  Choose $\gamma_0 := \max(\alpha_0,\beta_0)$.  Notice that
even $\liminf_\lambda \A \cap \liminf_\lambda \B =
      \liminf_{\alpha \to \lambda} (\A(\alpha) \cap \B(\alpha))$,
      since the reverse direction is follows immediately from
      $\A(\alpha) \cap \B(\alpha) \subseteq \A(\alpha),\B(\alpha)$.

For proposition~(4), assume $r \rhd r' \in \liminf_\lambda \A$. 
There exists $\alpha_0$ such that 
for all $\alpha \geq \alpha_0$, we have $r' \in A(\alpha)$, and thus,
$r \in \tri(\A(\alpha))$.  It follows that $r \in \liminf_{\alpha \to
  \lambda} \tri(\A(\alpha))$.
\qed

The last lemma can be dualized to $\limsup$:
\begin{lem}[$\limsup$ pushes through the building blocks of saturated sets]
  \label{lem:pushsat}
  \hfill\\
$
  \begin{array}{@{\ \ \quad}l@{~}l@{~}l@{~}l}
  (1) & \limsup_{\alpha \to \lambda} (r \sapp \A(\alpha)) & \subseteq & 
     r \sapp \limsup_\lambda \A.\\
  (2) & \limsup_{\alpha \to \lambda} e^{-1}(\A(\alpha)) & \subseteq &
     e^{-1}(\limsup_\lambda \A).\\
  (3) & \limsup_{\alpha \to \lambda} (\A(\alpha) \cup \B(\alpha)) & \subseteq &
     \limsup_\lambda \A \cup \limsup_\lambda \B. 
     \quad\mbox{(only classically!)}\\
  (4) & \limsup_{\alpha \to \lambda} \tri(\A(\alpha)) & \subseteq &
     \tri(\limsup_\lambda \A) \mbox{ if } \A(\alpha) \subseteq \V
     \mbox{ for all } \alpha < \lambda.\\
  \end{array}
$
\end{lem}
\proof
  The first three propositions follow trivially since the infimum and
  supremum considered are set-theoretic intersection and union.  
  Note that proposition~(3) is valid in classical logics but not
  in intuitionistic logics:
  An $r$ which inhabits infinitely many unions
  $\A(\alpha) \cup \B(\alpha)$, must classically inhabit infinitely
  many $\A(\alpha)$ or infinitely many $\B(\alpha)$.  But we cannot
  tell which of these alternatives holds, so the proposition has no
  intuitionistic proof.  However, we will not need this proposition
  for the results to follow.

  For the last proposition, assume $r \in \limsup_{\alpha \to \lambda}
  \tri(\A(\alpha))$.  If we require all following ordinals $< \lambda$,
  this means that for arbitrary $\alpha_0$
  there exist $\alpha \geq \alpha_0$ and $r' \in \A(\alpha)$ such that
  $r \rhd r'$.  Since each $r' \in \V$ and safe reduction into $\V$
  is deterministic, there is in fact a unique $r' \in \bigcup_{\alpha
    \geq \alpha_0} \A(\alpha)$ with $r \rhd r'$ for all $\alpha$,
  hence, $r \in \tri(\limsup_{\lambda} \A)$. 
\qed
\noindent
Proposition~(4) of Lemma~\ref{lem:pushsat} fails if we drop the
condition $\A(\alpha) \subseteq \V$: Define an infinite sequence $t_0,
t_1, \dots$ of terms by
$t_i = \tout^{i+1}(\tfixnu_0\,\tout)$ and observe that $t_i \rhd
t_{i+1}$.  Setting $\A(n) := \{t_i \mid i \geq n\}$ we have $t_0 \in
\inf_{n<\omega} \tri(\A(n)) \subseteq \limsup_{n \to \omega}
\tri(\A(n))$, but $\limsup_{\omega} \A = \inf_\omega \A = \emptyset$,
thus, $t_0 \not\in \tri(\limsup_{\omega} \A)$.  It did not help that
the $\A(n)$ were closed unter $\rhd$-reduction.

\noindent
\begin{lem}
  Binary sums $\bplus$ and products $\btimes$ and the operations
  $(-)^\mu$ and $(-)^\nu$ are $\limsup$-pushable
  and $\liminf$-pullable.
\end{lem}
\proof
Directly by the last lemmata.  For instance,
\[
\begin{array}{lll}
  (\liminf_\lambda \A) \btimes (\liminf_\lambda \B) 
  & = & (\tfst^{-1} \liminf_{\lambda} \A) \cap
        (\tsnd^{-1} \liminf_{\lambda} \B) \\
  & \subseteq & (\liminf_{\alpha\to\lambda} \tfst^{-1}\A(\alpha)) \cap
        (\liminf_{\beta\to\lambda}  \tsnd^{-1}\B(\beta)) \\
  & \subseteq & \liminf_{\gamma \to \lambda}
        (\tfst^{-1}\A(\gamma) \cap \tsnd^{-1}\B(\gamma)) \\
  & = & \liminf_{\gamma \to \lambda}
        (\A(\gamma) \btimes \B(\gamma)) . \\
\end{array}
\]
Because we wish to avoid classical reasoning
(Lemma~\ref{lem:pushsat}~(3)) as much as possible,
pushing $\limsup$ through disjoint sums requires a closer look:
Assume $r \in \limsup_{\gamma \to \lambda} (\A(\gamma) \bplus
\B(\gamma))$, hence, for some $\gamma$, either $r \rhd \tinl\, r'$
for some $r' \in \A(\gamma)$, or $r \rhd \tinr\, r'$ for some $r' \in
\B(\gamma)$, or $r \in \N$.  Since safe reduction $\rhd$ is
deterministic, and $\N$ does not contain values, one of these
three alternatives must hold whenever $r \in \A(\gamma) \bplus
\B(\gamma)$ for some $\gamma$.  So either $r \in \tri(\tinl\cdot(\limsup_\lambda
\A))$, or $r \in \tri(\tinr\cdot(\limsup_\lambda\B))$, or $r \in \N$, which means
$r \in (\limsup_\lambda \A) \bplus(\limsup_\lambda \B)$.

Analogously, we show that $\limsup$ pushes through $(\cdot)^\mu$.
\qed
\noindent
Using monotonicity of the product constructor, the lemma entails
that $\A(\alpha) \btimes \B(\alpha)$ is upper/lower semi-continuous if
$\A(\alpha)$ and $\B(\alpha)$ are.  This applies also for $\bplus$.

\begin{thm}[Pushing $\limsup$ through function space] \label{thm:limsupfun} \bla \\
\( \limsup_{\alpha \to \lambda}\ (\A(\alpha) \bto \B(\alpha))
    \subseteq 
   (\liminf_\lambda \A) \bto \limsup_\lambda \B .
\)
\end{thm}
\proof  We use Lemma~\ref{lem:fact}~(\ref{it:pushinfdep}).
\[
\def\arraystretch{1.3}
\begin{array}{lll}
  \limsup_{\alpha \to \lambda}\ (\A(\alpha) \bto \B(\alpha))
  & = & \limsup_{\alpha \to \lambda} 
          \bigcap_{s\in \A(\alpha)} (\_\,s)^{-1}\B(\alpha) \\
  & \subseteq & \bigcap_{s \in \liminf_\lambda \A} \limsup_{\alpha \to
    \lambda} (\_\,s)^{-1}\B(\alpha) \\
  & \subseteq & \bigcap_{s \in \liminf_\lambda \A} (\_\,s)^{-1}(\limsup_\lambda \B)
          \\
  & = & (\liminf_\lambda \A) \bto \limsup_\lambda \B.
\end{array}
\]
\qed
\begin{cor}\label{cor:fun}
  If $\A(\lambda) \lsub \liminf_\lambda \A$ and $\limsup_\lambda \B
  \lsub \B(\lambda)$ then $\limsup_\lambda (\A(\alpha) \bto
  \B(\alpha)) \lsub \A(\lambda) \bto \B(\lambda)$.
\end{cor}

\para{Coinductive types preserve upper semi-continuity} 

We have already seen that $\STREAM^\alpha(\NAT^\omega)$ is upper
semi-continuous.  In this section, we establish means to show that
also a type like $\STREAM^\alpha(\NAT^\alpha)$ is upper
semi-continuous (which is, by the way, isomorphic to $\NAT^\alpha \bto
\NAT^\alpha$).
\begin{defi}
A family $\F_\alpha \in \LL \to \LL'$ $(\alpha \in \O)$ is called
$\limsup$-pushable if for any $\G_{(-)} \in \O \to \LL$,
\[
\begin{array}{lll@{\qquad}l}
\limsup_{\alpha \to \lambda} \F_\gamma (\G_\alpha) & \lsub &
    \F_\gamma (\limsup_{\alpha \to \lambda} \G_\alpha)
& \mbox{for all } \gamma \in \O,
\\
\limsup_{\alpha \to \lambda} \F_\alpha (\G_\alpha) & \lsub &
    \F_\lambda (\limsup_{\alpha \to \lambda} \G_\alpha)
& \mbox{for all limits } \lambda > 0.  
\end{array}
\]  
\end{defi}
The first property is easier to prove, but not entailed by the second
property.  Usually we will confine in showing the second.
\begin{lem}\label{lem:pushnusecond}
  Let $\F_\alpha \in \vec\LL \to \LL \toco \LL$ be a family which is
  $\limsup$-pushable in all arguments..
Then for all $\beta$ the family
\[
\begin{array}{l}
  \H_\alpha \in \vec \LL \to \LL \\
  \H_\alpha(\vec \G) = \NU^\beta (\F_\alpha(\vec \G))
\end{array}
\]
is $\limsup$-pushable.
\end{lem}
\begin{proof}
  By transfinite induction on $\beta$ we prove
  for all $\G_i \in \O \to \LL_i$ that
\[
  \limsup_{\alpha \to \lambda} \NU^\beta (\F_\alpha(\vec \G_\alpha)) \lsub
  \NU^\beta \F_\lambda (\limsup_\lambda \vec\G).
\]
  In case $\beta = 0$, both sides become the
  maximum element of $\LL$.  In the successor case we have
\[
\begin{array}{l@{\hspace{1ex}}l@{\hspace{1ex}}l@{\hspace{2ex}}l}
 \limsup_{\alpha\to\lambda} \Inutwo{\beta+1}{(\F_\alpha(\vec \G_\alpha))} 
   & =  & \limsup_{\alpha\to\lambda} \F_\alpha(\vec \G_\alpha) 
   (\Inutwo\beta{(\F_\alpha(\vec \G_\alpha))}) \\
   & \lsub & \F_\lambda (\limsup_\lambda \vec \G)
   (\limsup_{\alpha\to\lambda} \Inutwo\beta{(\F_\alpha(\vec\G_\alpha))}) 
      & \F_\alpha \mbox{ pushable} \\
   & \lsub & \F_\lambda (\limsup_\lambda \vec \G)
   (\Inutwo\beta{(\F_\lambda(\limsup_\lambda \vec \G))}) 
      & \F_\alpha \mbox{ monotone, i.h.} \\
   & = & \Inutwo{\beta+1}{(\F_\lambda(\limsup_\lambda\vec \G))}. \\
\end{array}
\]
 In the remaining case $\beta = \lambda$ we exploit that
$\limsup$ pushes through infima (Lemma~\ref{lem:fact}.\ref{it:pushinf}).
\end{proof}

In the remainder of this part we will show that $\limsup_{\alpha <
  \lambda} \NU^{\phi(\alpha)} \F_\alpha \lsub \NU^{\liminf_\lambda \phi}
\F_\lambda$.  This will enable us to show that types like
$\STREAM^\alpha(\NAT^\alpha)$ are $\limsup$-pushable.

In the following, we will need additional properties of $\limsup$.
For the value of $\liminf_\lambda f$ and $\limsup_\lambda f$, only the
behavior of $f$ on a final segment of $[0;\lambda[$ is relevant:
\begin{lem}[Limit starting later] \label{lem:limlater} 
  Let ordinals range below $\lambda$. 
\[
\begin{array}{llll}
  (1) &  \inf_{\beta_0 \geq \alpha_0} \sup_{\beta \geq \beta_0} f(\beta)
  & = & \inf_{\gamma_0 \geq 0} \sup_{\gamma \geq \gamma_0} f(\gamma) .
\\
  (2) &  \sup_{\beta_0 \geq \alpha_0} \inf_{\beta \geq \beta_0} f(\beta)
  & = & \sup_{\gamma_0 \geq 0} \inf_{\gamma \geq \gamma_0} f(\gamma) .
\\
\end{array}
\]
\end{lem}
\begin{proof}
  We show $(1)$, the proof of $(2)$ is analogous.
  Direction $\lSuper$ follows by monotonicity of $\inf$. For
  $\lsub$, we have to show that for all $\gamma_0$ there exists a
  $\beta_0 \geq \alpha_0$ such that
\[ 
  \sup_{\beta \geq \beta_0} f(\beta)
  \lsub \sup_{\gamma \geq \gamma_0} f(\gamma) .
\]
  Take $\beta_0 = \max(\gamma_0,\alpha_0)$.
\end{proof}
\begin{lem}[Splitting limits]\label{lem:limsplit}
\[
\begin{array}{lll}
(1) & \limsup_{\beta \to \lambda} h(\beta,\beta) 
      \lsub
      \limsup_{\alpha \to \lambda} \limsup_{\beta \to \lambda} h(\alpha,\beta)
    & \mbox{for } h \in \O \tocontra \O \to \LL 
    ,
\\
(2) & \liminf_{\alpha \to \lambda} \liminf_{\beta \to \lambda} h(\alpha,\beta)
      \lsub 
      \liminf_{\beta \to \lambda} h(\beta,\beta)
    & \mbox{for } h \in \O \toco \O \to \LL
    .
\end{array}
\]
\end{lem}
\begin{proof}
  Again, we show just (1). Because of antitonicity, we have for all $\alpha$,
\[
\begin{array}{lll@{\hspace{2ex}}l}
  h(\beta,\beta) 
  & \lsub & h(\alpha,\beta) 
    & \mbox{for all } \beta \geq \alpha 
\\
  \sup_{\beta \geq \beta_0} h(\beta,\beta) 
  & \lsub &  \sup_{\beta \geq \beta_0} h(\alpha,\beta) 
    & \mbox{for all } \beta_0 \geq \alpha 
\\
  \inf_{\beta_0 \geq \alpha} \sup_{\beta \geq \beta_0} h(\beta,\beta) 
   & \lsub &    
  \inf_{\beta_0 \geq \alpha} \sup_{\beta \geq \beta_0} h(\alpha,\beta)
\\
  \limsup_{\beta \to \lambda} h(\beta,\beta)
   & \lsub &    
  \limsup_{\beta \to \lambda} h(\alpha,\beta)
   & \mbox{by Lemma~\ref{lem:limlater}} .
\end{array}
\]
  The goal follows by taking $\limsup$ on the r.h.s.
\end{proof}

Next, we show how to push a $\limsup$ into $\NU^{(-)}$.
\begin{lem}\label{lem:nuinfsup}
  Let $\phi \in \O \to \O$ and $I \subseteq \O$. Then
  \begin{enumerate}
  \item\label{it:one} $\sup_{\alpha\in I} \NU^{\phi(\alpha)} \lsub \NU^{\inf_I \phi}$,
  \item\label{it:two} $\sup_{\alpha\in I} \NU^{\phi(\alpha)} \lSuper \NU^{\inf_I \phi}$,
  \item\label{it:three} $\inf_{\alpha\in I} \NU^{\phi(\alpha)} \lSuper \NU^{\sup_I
      \phi}$, and
  \item\label{it:four} $\inf_{\alpha\in I} \NU^{\phi(\alpha)} \lsub \NU^{\sup_I \phi}$.
  \end{enumerate}
\end{lem}
\begin{proof}
  (\ref{it:one}) and (\ref{it:three}) follow directly from
  antitonicity.  For (\ref{it:two}), remember that each set of ordinals
  is left-closed, hence $\inf_I \phi = \phi(\alpha)$ for some $\alpha
  \in I$.  The remaining proposition (\ref{it:four}) is proven by cases
  on $\sup_I \phi$.  If $\sup_I \phi$ is not a limit ordinal then
  $\sup_I \phi = \phi(\alpha)$ for some $\alpha \in I$.  For this
  $\alpha$, clearly $\NU^{\phi(\alpha)} \lsub \NU^{\sup_I \phi}$,
  which entails the lemma.  Otherwise, if $\sup_I \phi$ is a limit
  ordinal, then by definition of $\NU$ at limits we have to show
  $\inf_{\alpha\in I} \NU^{\phi(\alpha)} \lsub \NU^{\beta}$ for all
  $\beta < \sup_I \phi$.  By definition of the supremum, $\beta <
  \phi(\alpha)$ for some $\alpha$.  Since $\NU$ is antitone,
  $\NU^{\phi(\alpha)} \lsub \NU^\beta$ from which we infer our subgoal
  by forming the infimum on the left hand side.
\end{proof}
\begin{cor}\label{cor:pushnufirst}
  $\limsup_{\alpha\to\lambda} \NU^{\phi(\alpha)} =
      \NU^{\liminf_\lambda \phi}$.
\end{cor}

Now we have the tools in hand to prove that coinductive types preserve
upper semi-continuity.  In the second part of the following theorem,
we make use of the fact that our coinductive types close at ordinal $\omega$.
\begin{thm}[Upper semi-continuity of coinductive types]
    \label{thm:nucont}
  Let $\F_\alpha \in \vec\LL \to \LL \toco \LL$ be a family which is
  $\limsup$-pushable in all arguments
  and $\phi \in \O \toco \O$. Then
\[
  \limsup_{\alpha\to\lambda} \NU^{\phi(\alpha)} (\F_\alpha(\vec \G_\alpha))
  \lsub \NU^{(\liminf_\lambda \phi)}(\F_\lambda(\limsup_\lambda\vec\G)).
\]
  If $\phi$ is affine, then even
\[
  \limsup_{\alpha\to\lambda} \NU^{\phi(\alpha)} (\F_\alpha(\vec \G_\alpha))
  \lsub \NU^{\phi(\lambda)}(\F_\lambda(\limsup_\lambda\vec\G)).
\]
\end{thm}
For our purposes, an \emph{affine} function on $\O$ takes the shape
$\phi(\alpha) = \min \{ b\alpha + \beta, \ltop[\tord] \}$ for some $b \in
\{0,1\}$ and $\beta \in \O$.
\begin{proof}
  Direct.  Note that $\NU^{\phi(-)}$ is antitone, so we can split the $\limsup$.
\[
 \begin{array}{lll@{\hspace{3ex}}l}
   \limsup_{\alpha\to\lambda} \NU^{\phi(\alpha)} (\F_\alpha(\vec\G_\alpha))
 & \lsub & \limsup_{\alpha\to\lambda} \limsup_{\gamma\to\lambda}
    \NU^{\phi(\alpha)} (\F_\gamma(\vec\G_\gamma))
    & \mbox{Lemma~\ref{lem:limsplit}} \\
 & \lsub & \limsup_{\alpha\to\lambda} \NU^{\phi(\alpha)}
     (\F_\lambda(\limsup_\lambda \vec \G))
    & \mbox{Lemma~\ref{lem:pushnusecond}} \\
 & \lsub & \NU^{\liminf_{\alpha\to\lambda}\phi(\alpha)}
     (\F_\lambda(\limsup_\lambda \vec \G))
    & \mbox{Cor.~\ref{cor:pushnufirst}} .
 \end{array}
\]
  Now we consider affine $\phi$. 
  If $\phi$ is constant, then $\liminf_\lambda \phi =
  \phi(\lambda)$.  Otherwise, $\phi(\alpha) \geq \alpha$, hence, $\phi(\lambda)
  \geq \omega$, and also $\liminf_\lambda \phi \geq \omega$.  We only
  need to show that in our case the greatest fixed-point is
  reached already at iteration $\omega$.  
  Observe that $\H(\X) := \F_\lambda(\limsup_\lambda \vec \G)(\X)$ is
  $\limsup$-pushable, since the family $\F$ is.  It suffices to show that
  $\NU^\omega \H \lsub \NU^{\omega+1} \H$.
\[
\begin{array}{lclclcl}
  \NU^\omega \H 
  & = & \limsup_{\beta \to \omega} \NU^\beta \H 
  & = & \limsup_{\beta \to \omega} \NU^{\beta+1} \H
  & = & \limsup_{\beta \to \omega} \H(\NU^\beta \H) \\
  & \lsub & \H(\limsup_{\beta \to \omega} \NU^\beta \H) 
  & = & \NU^{\omega+1} \H . \\
\end{array}
\]
  Thus, $\NU^{\omega+1} \H = \NU^\omega \H$, which means that
  $\NU^\beta \H = \NU^\omega \H$ for all $\beta \geq \omega$; the
  fixed-point is reached.
\end{proof}
For example, since $\F_\alpha(\X) = (\NAT^\alpha \btimes \X)^\nu$ is
$\limsup$-pushable, we have can infer upper semi-continuity of
$\STREAM^\alpha(\NAT^\alpha) = \NU^\alpha \F_\alpha$.

\para{Inductive types preserve lower semi-continuity.}
We can dualize the results of the last section and prove that
inductive types preserve lower semi-continuity and
$\limsup$-pushability.
\begin{defi}
  A family $\F_\alpha \in \LL \to \LL'$ $(\alpha \in \O)$ is called
  $\liminf$-pullable if for all $\G_{(-)} \in \O \to \LL$,
\[
\begin{array}{lll@{\qquad}l}
  \F_\gamma(\liminf_{\alpha \to \lambda} \G_\alpha) & \lsub &
  \liminf_{\alpha\to\lambda} \F_\gamma(\G_\alpha)
& \mbox{for all } \gamma \in \O,
\\
  \F_\lambda(\liminf_{\alpha \to \lambda} \G_\alpha) & \lsub &
  \liminf_{\alpha\to\lambda} \F_\alpha(\G_\alpha)
& \mbox{for all limits } \lambda > 0.
\end{array}
\]
\end{defi}
\begin{lem}
  Let $\F_\alpha \in \vec \LL \to \LL \toco \LL$ be a family which is
  $\liminf$-pullable in all arguments. Then,
  \begin{enumerate}
  \item $\MU^\beta (\F_{(-)}(-))$ is a $\liminf$-pullable family,
  \item $\MU^{(\liminf_\lambda \phi)} = 
         \liminf_{\alpha \to \lambda} \MU^{\phi(\alpha)}$.
  \end{enumerate}
\end{lem}
\begin{thm}[Lower semi-continuity of inductive types]
    \label{thm:mucont}
  Let $\F_\alpha \in \vec\LL \to \LL \toco \LL$ be a family which is
  $\liminf$-pullable in all arguments
  and $\phi \in \O \toco \O$. Then
\[
   \MU^{(\liminf_\lambda \phi)} (\F_\lambda(\liminf_\lambda \vec \G))
         \lsub 
         \liminf_{\alpha \to \lambda} \MU^{\phi(\alpha)}
         (\F_\alpha(\vec \G_\alpha))
   .
\]
  If $\phi$ is lower semi-continuous, then even
\[
   \MU^{\phi(\lambda)} (\F_\lambda(\liminf_\lambda \vec \G))
         \lsub 
         \liminf_{\alpha \to \lambda} \MU^{\phi(\alpha)}
         (\F_\alpha(\vec \G_\alpha))
   .
\]
\end{thm}
Putting together the conditions $\phi$ is required to be monotone and
continuous in the second statement of the theorem.  Since $\phi$ is
coming from a size expression in our case, such a $\phi$ will either
be the identity or a constant function.  (The successor function is
\emph{not} continuous!)

Using Thm.~\ref{thm:mucont}, we can establish lower semi-continuity of
$\LIST^\alpha(\NAT^\alpha)$.

\section{Non Semi-Continuous Types}
\label{sec:negres}

This section is devoted to show that our list of criteria for
semi-continuity is somewhat complete.  Concretely, we demonstrate
that there is no compositional scheme to establish lower
semi-continuity of function types or upper semi-continuity of inductive
types.

\para{Function space and lower semi-continuity} One may wonder
whether Cor.~\ref{cor:fun} can be dualized, \ie, does upper
semi-continuity of $\A$ and lower semi-continuity of $\B$ entail lower
semi-continuity of $\C(\alpha) = \A(\alpha) \bto \B(\alpha)$?
The answer is no, \eg, consider $\C(\alpha)=\NAT^\omega \bto
\NAT^\alpha$.  Although $\A(\alpha)=\NAT^\omega$ is trivially upper
semi-continuous, and $\B(\alpha)=\NAT^\alpha$ is lower
semi-continuous, $\C$ is not lower semi-continuous:  For instance, the
identity function is in $\C(\omega)$ but in no $\C(\alpha)$ for
$\alpha<\omega$, hence, also not in $\liminf_\omega \C$.
And indeed, if this $\C$ was lower semi-continuous, then our criterion
would be unsound, because then by Cor.~\ref{cor:fun} the type
$(\NAT^\omega \bto \NAT^\alpha) \bto \NAT^\omega$, which admits a
looping function (see introduction), would be upper
semi-continuous.

Nevertheless, there are some lower semi-continuous function spaces,
for instance, if the domain is a finite type.  For example, $\BOOL \bto
\NAT^\alpha$ is lower semi-continuous in $\alpha$, which implies that
$(\BOOL \bto \NAT^\alpha) \bto \NAT^\alpha$ could be admissible. This
is the type of a maximum function taking its two inputs in form of a
function over booleans ($\BOOL \bto \NAT^\alpha \cong \NAT^\alpha
\btimes \NAT^\alpha$).  However, this example is somewhat contrived;
it is not clear whether such cases appear in practice, so we will not
pursue this further here.

\para{Inductive types and upper semi-continuity}  Pareto
\cite{pareto:PhD} proves that inductive types are (in our terminology)
$\limsup$-pushable.  His inductive types denote only finitely
branching trees, but we also consider infinite branching, arising from
function space embedded in inductive types.  
\shortversion{%
In my thesis \cite[Sect.~5.4.3]{abel:PhD}
I show that infinitely branching inductive data types do not inherit
upper semi-continuity from their defining body.
But remember that
inductive types can still be upper semi-continuous if they are
covariant in their size index.
}%
\longversion{%
Such an infinitely branching type is the
type of hungry functions (which consumes one argument after the other):
\[
\begin{array}{lll}
  \Hungry & : & \tord \toco \ttype \tocontra \ttype \\
  \Hungry & := & \lambda \vari \lambda A.\ \mu^\vari \lambda X.\, A
  \to X .\\
\end{array}
\]
We are interested in the special case of $\Hungry^\vari(\Nat^\vari)$.
In the following we show that accepting such a type as the result of
recursion will lead to accepting a non-terminating program.  As a
consequence, infinitely branching inductive data types, such as
$\mu^\vari \lambda X.\, \Nat^\vari \to X$, do not inherit
upper semi-continuity from their defining body, here, $\Nat^\vari \to
X$ (recall that $\Nat^\vari$ is lower semi-continuous, hence
$\Nat^\vari \to X$ is upper semi-continuous).  But remember that
inductive types can still be upper semi-continuous, \eg, 
$\Hungry^\vari(\Nat^\infty) = \mu^\vari \lambda X.\, \Nat^\infty \to
X$, which is covariant in its size index.

Semantically, we set $\H^\alpha = \MU^\alpha \F_\alpha$, where
$\F_\alpha(\X) = (\NAT^\alpha \bto \X)^\mu$.  Since 
\[
\begin{array}{rcl}
\limsup_{\alpha \to\lambda} \F_\alpha(\G(\alpha)) 
& \lsub & 
  ((\liminf_{\alpha\to\lambda} \NAT^\alpha) \bto (\limsup_\lambda \G))^\mu 
\\
& = & 
  \F_\lambda(\limsup_\lambda \G),
\end{array}
\] 
the family $\F_\alpha$ is $\limsup$ pushable.  If we had a result like
\[
  \limsup_{\alpha\to\lambda} \Imutwo{\phi(\alpha)}{\F_\alpha} \lsub
    \Imutwo{\limsup_\lambda\phi}{\limsup_\lambda\F},
\]
then $\H$ would be upper semi-continuous, and it would be legal to
write the following recursive function:
\[
\begin{array}{@{\hspace{0ex}}llllll@{\hspace{0ex}}}
  \thu & : & \multicolumn{3}{@{\hspace{0ex}}l@{\hspace{0ex}}}{
     \forall \vari.\, \Nat^{\vari} \to \Hungry^\vari(\Nat^{\vari}) 
  }\\
  \thu & := & \multicolumn{3}{@{\hspace{0ex}}l@{\hspace{0ex}}}{
     \tfixmu_0 \lambda h\lambda \_.\ \tin\,(\tss\circ h \circ \tpred) 
  }\\[1ex]
  \multicolumn{5}{@{\hspace{0ex}}l@{\hspace{0ex}}}{
    \thu\ (\tin\,\_) \ \red^+ \tin\,(\tss\circ \thu \circ \tpred) }\\
\end{array}
\]
We will show that the existence of $\thu$ destroys normalization.
In the body of $\thu$ we refer to an auxiliary function $\tss$.  As
well as its inverse, $\tp$, it can be defined by induction on $\vari$: 
\[
\begin{array}[t]{@{\hspace{0ex}}llllll@{\hspace{0ex}}}
  \tss& : & \multicolumn{3}{@{\hspace{0ex}}l@{\hspace{0ex}}}{
     \forall \vari \forall \varii.\, \Hungry^\vari(\Nat^{\varii}) \to
     \Hungry^\vari(\Nat^{\varii+1}) 
  }\\
  \tss& := & \multicolumn{3}{@{\hspace{0ex}}l@{\hspace{0ex}}}{
     \tfixmu_0 \lambda s\lambda h.\ \tin\,(s \circ (\tout\,h) \circ \tpred) 
  }\\[1ex]
  \multicolumn{5}{@{\hspace{0ex}}l@{\hspace{0ex}}}{
    \tss\ (\tin\,f) \red^+  \tin\,(\tss \circ f \circ \tpred) } 
\\[2ex]
  \tp & : & \multicolumn{3}{@{\hspace{0ex}}l@{\hspace{0ex}}}{
     \forall \vari \forall \varii.\, \Hungry^\vari(\Nat^{\varii+1}) \to
     \Hungry^\vari(\Nat^{\varii}) 
  }\\
  \tp & := & \multicolumn{3}{@{\hspace{0ex}}l@{\hspace{0ex}}}{
     \tfixmu_0\lambda p\lambda h.\ \tin\, (p \circ (\tout\,h) \circ \tsucc)
  }\\[1ex]
  \multicolumn{5}{@{\hspace{0ex}}l@{\hspace{0ex}}}{
    \tp\ (\tin\,f)  \red^+  \tin\,(\tp \circ f \circ \tsucc) } \\
\end{array}
\]
(Note that these definitions are perfectly acceptable and not to blame.)
Another innocent function is the following.  Its type looks funny,
since it produces something in the empty type, but let us mind that
$\Hungry$, being an inductive type ``with nothing to start
induction,'' is also empty.
\[
\begin{array}{@{\hspace{0ex}}llllll@{\hspace{0ex}}}
  \ttr & : & \multicolumn{3}{@{\hspace{0ex}}l@{\hspace{0ex}}}{
     \forall\vari.\, \Hungry^\vari(\Nat^{\infty}) \to \forall A. A
  }\\
  \ttr & := & \multicolumn{3}{@{\hspace{0ex}}l@{\hspace{0ex}}}{
     \tfixmu_0 \lambda \vtr\lambda h.\ \vtr\ ((p \circ (\tout\,h) \circ \tsucc)\ \tzero) 
  }\\[1ex]
  \multicolumn{5}{@{\hspace{0ex}}l@{\hspace{0ex}}}{
    \ttr\ (\tin\,f)  \red^+  \ttr\ ((\tp \circ f \circ \tsucc)\ \tzero) } \\
\end{array}
\]
Some calculation now shows that $\ttr\,(\thu\,\tzero)$, the application of
$\ttr$ to the ``bad guy'' $\th$, diverges:
\[
\begin{array}{llll}
\ttr\,(\thu\,\tzero)
& \red^+ & \ttr\,(\tin\,(\tss \circ \thu \circ \tpred)) \\
& \red^+ & \ttr\,((\tp \circ \tss \circ \thu \circ \tpred \circ
  \tsucc)\,\tzero) & \red^+ \\
\ttr\,(\tp\,(\tss\,(\thu\,\tzero))) 
& \red^+ & \ttr\,(\tp\,(\tss\,(\tin\,(\tss \circ \thu \circ
\tpred))))) \\
& \red^+  & \ttr\,(\tp\,(\tin\,(\tss^2 \circ \thu \circ \tpred^2))) \\
& \red^+  & \ttr\,(\tin\,(\tp \circ \tss^2 \circ \thu \circ \tpred^2
\circ \tsucc)) \\
& \red^+  & \ttr\,((\tp^2 \circ \tss^2 \circ \thu \circ \tpred^2
\circ \tsucc^2)\,\tzero) & \red^+  \\ 
\ttr\,(\tp^2\,(\tss^2\,(\thu\,\tzero))) & \red^+  & \dots \\
\end{array}
\]
}

\begin{deffigure}{\textwidth}{\Fhat: Semi-continuous constructors and
    recursion types.\label{fig:semi}}
Strictly positive contexts.
\[
\begin{array}{lrl@{\hspace{4ex}}l}
 \Pi & ::= & \cempty \mid \Pi, X\ofkind{\pco\pkind} 
\end{array}
\]
Semi-continuity $\Delta \sep \Pi \derq F : \kappa$ for $q \in\{\pupper,\plower\}$.
\begin{gather*}
  \nru{\rcontco
     }{\Delta, \vari \of \pco\tord \derD F : \kappa \qquad p \leq \pco
     }{\Delta, \vari \of p\tord \sep \Pi \deru F : \kappa}
\qquad
  \nru{\rcontcontra
     }{\Delta, \vari \of \pcontra\tord \derD F : \kappa \qquad p \leq \pcontra
     }{\Delta, \vari \of p\tord \sep \Pi \derl F : \kappa}
\\[2ex]
  \nru{\rcontin
     }{\Delta \derD F : \kappa
     }{\Delta, \vari \of p\tord \sep \Pi \derq F : \kappa}
\qquad
  \nru{\rcontvar
    }{X\ofkind{p\kappa} \in \Delta,\Pi \qquad p \leq \pco 
    }{\Delta \sep \Pi \derq X : \kappa}
\\[2ex]
  \nrux{\rcontabs
    }{\Delta, X \ofkind{p\kappa} \sep \Pi \derq F : \kappa'
    }{\Delta \sep \Pi \derq \lambda X\ofkinddot\kappa F : p\kappa \to \kappa'
    }{X \not= \vari}
\\[2ex]
  \nru{\rcontapp
    }{\Delta, \vari\of p'\tord \sep \Pi \derq F : p\kappa \to \kappa' \qquad
      \pinv p\Delta \derD G : \kappa
    }{\Delta, \vari\of p'\tord \sep \Pi \derq F\,G : \kappa'  }
\\[2ex]
  \nru{\rcontsum
     }{\Delta \sep \Pi \derq A : \ttype \qquad
       \Delta \sep \Pi \derq B : \ttype 
     }{\Delta \sep \Pi \derq A + B : \ttype}
\\[2ex]
  \nru{\rcontprod
     }{\Delta \sep \Pi \derq A : \ttype  \qquad
       \Delta \sep \Pi \derq B : \ttype
     }{\Delta \sep \Pi \derq A \times B : \ttype}
\\[2ex]
  \nru{\rcontarr
     }{-\Delta \sep \cempty \derl A : \ttype \qquad
        \Delta \sep \Pi \deru B : \ttype  
     }{\Delta \sep \Pi  \deru A \to B : \ttype}
\qquad
  \nru{\rcontforall
     }{\Delta \sep \Pi \deru F : \pin\kappa \to \ttype
     }{\Delta \sep \Pi \deru \forallk\kappa F : \ttype}
\\[2ex]
  \nrux{\rcontmu
      }{\Delta \sep \Pi, X\of \pco \pkind \derl F : \pkind \qquad 
        \Delta \der a : \tord 
      }{\Delta \sep \Pi \derl \mu^a (\lambda X\ofkinddot\pkind F) : \pkind
      }{a = \vari \mbox{ or } \vari \not\in\FV(a)}
\\[2ex]
  \nru{\rcontnu
     }{\Delta \sep \Pi, X\of \pco \pkind \deru F : \pkind \qquad 
       \Delta \der a \jord
     }{\Delta \sep \Pi \deru \nu^a (\lambda X\ofkinddot\pkind F) : \pkind}
\end{gather*}
Pure ordinal expressions $\Delta \der a \jord$.
\begin{gather*}
  \nru{\rordinf}{}{\Delta \der \infty \jord}
\quad
  \nru{\rordvar
     }{(\vari \of p\tord) \in \Delta \qquad p \leq +
     }{\Delta \der \vari \jord}
\quad
  \nru{\rords
     }{\Delta \der a \jord
     }{\Delta \der \s\,a \jord}
\end{gather*}
\end{deffigure}

\section{A Kinding System for Semi-Continuity}
\label{sec:der}

We turn the results of Section~\ref{sec:semicont} into a calculus and define a
judgement $\Delta\sep \Pi \derq F : \kappa$, where $\vari$ is an
ordinal variable with $(\vari \of p\tord) \in \Delta$ for some $p$, 
the bit $q \in \{ \plower,
\pupper \}$ states whether the constructor $F$ under consideration is
lower ($\plower$) or upper ($\pupper$) semi-continuous, and $\Pi$ is a
context of \emph{strictly positive} constructor variables $X\of \pco\kappa'$.
We will prove (Thm.~\ref{thm:soundcont}) that the family $F(\vari)$ is
$\limsup$-pushable in all $X \in \Pi$ if $q = \pupper$ and
$\liminf$-pullable if $q = \plower$.

The complete listing of rules can be found in
Figure~\ref{fig:semi}; in the following, we discuss
a few.
\[
  \nru{\rcontco
     }{\Delta, \vari \of \pco\tord \derD F : \kappa 
       \qquad p \in \{ \pco, \pin \}
     }{\Delta, \vari \of p\tord \sep \Pi \deru F : \kappa}
\]
If $\vari$ appears only positively in $F$, then $F$ is trivially upper
semi-continuous. However, monotonicity does not imply
$\limsup$-pushability, so no variables from $\Pi$ may occur in $F$.
In the conclusion we may choose to set $p=\pin$,
meaning that we forget that $F$ is monotone in $\vari$.
Rule $\rcontcontra$ is analogous and rule $\rcontin$ states that a
constant $F$ is (trivially) continuous.
\[
  \nru{\rcontvar
    }{X\ofkind{p\kappa} \in \Delta,\Pi \qquad p \leq \pco 
    }{\Delta \sep \Pi \derq X : \kappa}
\]
Rule $\rcontvar$ can be used also for $X = \iota$.  It states that the
identity is continuous and both $\limsup$-pushable and
$\liminf$-pullable.  

Using the four rules discussed so far, we can derive semi-continuity
properties of ordinal expressions.  Expressions like $\infty$ and
$\s^n\varii$ (with $\varii \not= \vari$)
which are constant in $\iota$ are trivially continuous;
so is the identity $\iota$.  Expressions of the form $\s^n\vari$ with
$n \geq 1$ are only upper semi-continuous (rule $\rcontco$), but not
lower semi-continuous.

Now we discuss some rules to construct semi-continuous types and type
constructors. 
\[
  \nru{\rcontarr
     }{\pcontra\Delta \sep \cempty \derl A : \ttype \qquad
        \Delta \sep \Pi \deru B : \ttype  
     }{\Delta \sep \Pi  \deru A \to B : \ttype}
\]
This rule incarnates Thm.~\ref{thm:limsupfun}.  Note that, because $A$ is
to the left of the arrow, the polarity of
all ordinary variables in $A$ is reversed, and $A$ may not
contain strictly positive variables.
\[
  \nru{\rcontnu
     }{\Delta \sep \Pi, X\of \pco \pkind \deru F : \pkind
       \qquad \Delta \der a \jord
     }{\Delta \sep \Pi \deru \nu^a (\lambda X\ofkinddot\pkind F) : \pkind}
\]
Rule $\rcontnu$ states that strictly positive coinductive types are
upper semi-continuous.  
The ordinal $a$ must be $\infty$ or $\s^n \varii$ for some
$\varii\ofkind\tord \in \Delta$ (which may also be identical to
$\vari$).
\begin{lem}
  \label{lem:semjord}
  Assume $\Delta \der a \jord$ and let $\theta \in \den\Delta$,
   $\vari$ an ordinal variable, and 
  \[\phi := (\alpha \mapsto \Den a
  {\update\theta\vari\alpha}) \in \den\tord \to \den\tord .\]
  Then $\phi$ is affine.
\end{lem}
\begin{proof}
  By induction on $\Delta \der a \jord$.
\end{proof}

\begin{thm}[Soundness of Continuity Derivations]\label{thm:soundcont}
  Assume $\Delta \sep \vec X \of \pco\vec\kappa \derq F : \kappa$.
  Let $\theta$ a valuation of the variables in $\Delta$
  and set $\F_\alpha(\vec \G) = \Den F{\update{\update\theta\vari\alpha} 
      {\vec X}{\vec \G}}$. 
  \begin{enumerate}

  \item If $q = \plower$ then the family $\F$ is $\liminf$-pullable. 

  \item If $q = \pupper$ then the family $\F$ is $\limsup$-pushable.
  \end{enumerate}
\end{thm}
\begin{proof}
  By induction on the continuity derivation.  Some cases:
\[
  \nru{\rcontnu
     }{\Delta \sep \vec X \of \vec \kappa, X\of \pco \pkind \deru F : \pkind
       \qquad \Delta \der a \jord
     }{\Delta \sep \vec X \of \vec \kappa \deru \nu^a (\lambda X\ofkinddot\pkind F) : \pkind}
\]
  Let $\F_\alpha(\vec \G)(\H) = \Den F{\update{\update{\update\theta\vari\alpha} 
      {\vec X}{\vec \G}}X\H}$ and $\phi(\alpha) = \Den a
  {\update\theta\vari\alpha}$.  By Lemma~\ref{lem:semjord}, $\phi$ is
  affine, and by induction hypothesis, $\F$ is
  $\limsup$-pushable.  Thus, we can apply Thm.~\ref{thm:nucont} to
  infer the goal.
\[
  \nrux{\rcontmu
      }{\Delta \sep \Pi, X\of \pco \pkind \derl F : \pkind \qquad 
        \Delta \der a : \tord 
      }{\Delta \sep \Pi \derl \mu^a (\lambda X\ofkinddot\pkind F) : \pkind
      }{a = \vari \mbox{ or } \vari \not\in\FV(a)}
\]
  $\phi(\alpha) := \Den a
  {\update\theta\vari\alpha}$ is either constant or the identity,
  hence, it is monotone and continuous.  The goal follows from
  Thm.~\ref{thm:mucont}. 
\end{proof}

Now we are able to formulate the syntactical admissibility criterion
for types of (co)re\-cur\-sive functions.
\begin{defi}[Syntactic admissibility]\label{def:synadm}
\[
\begin{array}{lll}
 \Gamma \der A \adm{\tfixmu_n} 
  & \mbox{iff} &
    \Gamma,\, \vari\ofkind{\pin\tord} \derD A\,\vari 
      = \forall \vec X\of\vec\kappa. 
        B_1 \to \dots \to B_n \to \mu^\vari F \vec H \to C: \ttype \\
  & \mbox{and} &
    \Gamma, \vari\ofkind{\pin\tord} ;\cempty \deru 
      \forall \vec X\of\vec\kappa.\, 
         B_{1..n} \to \mu^\vari F \vec H \to C : \ttype 
\\[2ex]
 \Gamma \der A \adm{\tfixnu_n} 
  & \mbox{iff} &
    \Gamma,\, \vari\ofkind{\pin\tord} \derD A\,\vari 
      = \forall \vec X\of\vec\kappa. 
        B_1 \to \dots \to B_n \to \nu^\vari F \vec H : \ttype \\
  & \mbox{and} &
    \Gamma, \vari\ofkind{\pin\tord};\cempty \deru
      \forall \vec X\of\vec\kappa.\, 
        B_{1..n} \to \nu^\vari F \vec H  : \ttype 
\end{array}
\]
\end{defi}
It is easy to check that admissible types fulfill the semantic
criteria given at the end of Section~\ref{sec:sem}.  We prove
Assumption~\ref{asm:semadm}, restated as the following theorem.
\begin{thm}[Soundness of admissibility]\label{thm:semadm}
If\/ $\Gamma \der A \adm{\tfixnabla_n\!}$ and $\theta(X) \in \den{\kappa}$
for all $(X\of \kappa) \in \den\Gamma$ then $\A := \Den A \theta \in
\den\tord \to \den\ttype$ has the
following properties:
\begin{enumerate}
\item%
 Shape: $\A(\alpha) = \bigcap_{k\in K} \B_1(k,\alpha) \bto \dots
\bto \B_n(k,\alpha) \bto \B(k,\alpha)$ for some $K$ and some $\B_1, \dots, \B_n,
\B \in K \times \den\tord \to \den\ttype$.  In case $\nabla=\mu$, $\B(k,\alpha) = \calI(k,\alpha)^\mu \bto
\C(k,\alpha)$ for some $\calI, \C$.  Otherwise, $\B(k,\alpha) =
\C(k,\alpha)^\nu$ for some $\C$.
\item%
 Bottom-check: $\calI(k,0)^\mu = \lbot[\ttype]$ in case $\nabla=\mu$
  and $\C(k,0)^\nu = \ltop[\ttype]$ in case $\nabla=\nu$.
\item%
 Limit-check: $\inf_{\alpha < \lambda} \A(\alpha) \subseteq \A(\lambda)$
  for all limit ordinals $\lambda \in \den\tord\setminus\{0\}$.
\end{enumerate}  
\end{thm}
\begin{proof}
  Set $K := \den{\kappa_1} \times \dots \den{\kappa_m}$ with $m :=
  |\vec\kappa|$ and $\B_i(k,\alpha) :=
  \Den{B_i}{\update{\update\theta{\vec X}k}\vari\alpha}$ for $i = 1..n$.
  Further, let $\F(k)(\alpha) := \Den F {\update{\update\theta{\vec
      X}k}\vari\alpha}$ and $\H_j(k)(\alpha) := \Den {H_j}
  {\update{\update\theta{\vec X}k}\vari\alpha}$.

  In case $\nabla = \mu$, first let $\C(k,\alpha) = \Den {C}
  {\update{\update\theta{\vec X}k}\vari\alpha}$.
  Define $\emptyset^{\vec p\vec\kappa \to \ttype}(\vec\G) = \emptyset$
  and observe that for any $f \in \den\kappa \toco \den\kappa$,
  $\MU^\alpha f^\mu = (f^\mu)^\alpha(\lbot[\kappa])
  = ((f \circ (-)^\mu)^\alpha(\emptyset^\kappa))^\mu$.  
  (Induction
  on $\alpha$, using $\emptyset^\mu = \lbot$ and $\sup$-continuity of
  $(-)^\mu$.)  Thus, we can set 
\[ \calI(k,\alpha) = (((\F(k)(\alpha) \circ
  (-)^\mu)^\alpha(\emptyset))(\vec \H)
\] and have $\calI(k,\alpha)^\mu =
  \Den {\mu^\vari F\,\vec H}{\update{\update\theta{\vec
        X}k}\vari\alpha}$.
  Properties (\ref{it:shape}) and (\ref{it:bottom}) are hence
  satisfied.  By Thm.~\ref{thm:soundcont}, the type $\A$ is upper
  semi-continuous which implies (\ref{it:limit}).

  For case $\nabla = \nu$, define $(\tout\cdot\ltop[\vec p\vec\kappa \to
  \ttype])(\vec \G) = \tout\cdot\ltop[\ttype]$.  Then
  $(\tout\cdot\ltop)^\nu = \ltop$.
  Observe that $\NU^\alpha f^\nu =
  (f^\nu)^\alpha(\ltop) =  ((f \circ (-)^\nu)^\alpha(\tout\cdot\ltop))^\nu$
  and set 
\[
  \C(k,\alpha) =  (((\F(k)(\alpha) \circ
  (-)^\nu)^\alpha(\tout\cdot\ltop))(\vec \H) .
\] 
  Then $\C(k,\alpha)^\nu = \Den {\nu^\vari F\,\vec H}
  {\update{\update\theta{\vec X}k}\vari\alpha}$. Using
  Thm.~\ref{thm:soundcont}, all three properties hold.
\end{proof}

\begin{exa}[Inductive type inside coinductive type]
  Rule $\rcontnu$ allows the type system to accept the following
  definition, which assigns an informative type to the stream $\tnats$ of
  all natural numbers in ascending order:
  \[
  \begin{array}{lll}
     \tnats & : & \forall \vari.\, \Stream^\vari\, \Nat^\vari \\
     \tnats & := & \tfixnu_0 \lambda \vnats.\, \ttuple{\tzero,\
       \tmapStream\ \tsucc\ \vnats} \\[1ex]
     \tmapStream & : & \forall A \forall B.\, (A\to B) \to 
        \forall \vari.\, \Stream^\vari A \to \Stream^\vari B 
     \\
     \tmapStream & := & \lambda f.\, \tfixnu_1 \lambda \vmaps \lambda
     s.\, \tin \ttuple{f\,(\tfst\,(\tout\,s)),\;
       \vmaps\,(\tsnd\,(\tout\,s))}
  \end{array}
  \]
  The type of $\tnats$ expresses that if you read the first $n$
  elements of the stream, these are numbers $< n$.  In particular,
  the $i$th element of $\tnats$ is at most $i-1$.  This is the most
  information a type of $\tnats$ can carry in our type system.
\end{exa}

\begin{exa}[Inductive type inside inductive type] \label{ex:bf}
In the following, we describe breadth-first traversal of rose (finitely
branching) trees whose termination is recognized by $\Fhat$.
\[
\begin{array}{lll}
\Rose & : & \tord \toco \ttype \toco \ttype \\
\Rose & := & \lambda \vari\lambda A.\,\GRose^\vari\,\List^\infty\,A
=  \lambda \vari\lambda A.\ \mu_\ttype^\vari \lambda X.\, 
   A \times \List^\infty X 
\end{array}   
\]
The step function, defined by induction on $\varii$, traverses a list
of rose trees of height $< \vari + 1$ and produces a list of the roots and
a list of the branches (height $< \vari$).
\[
\begin{array}{lll}
\tstep & : &  \forall \varii\forall A\forall \vari.\, 
  \List^\varii(\Rose^{\vari+1}\,A) \to 
  \List^\varii\,A \times \List^\infty (\Rose^\vari\,A) 
\\[1ex]
\tstep & := & \tfixmu_0 \lambda \vstep\lambda l.\, 
  \tmatchraw\;l\;\twith \\
&& \quad 
\begin{array}{l}
  \tnil \tmarr \ttuple{\tnil,\,\tnil} \\
  \tcons\,\ttuple{a,\vrs'}\,\vrs  \tmarr  
  \tmatchraw\;\vstep\,\vrs\;\twith \\
  \qquad \qquad
  \ttuple{\vas,\,\vrs''} \tmarr \ttuple{\tcons\;a\,\vas,\
    \tappend\,\vrs'\,\vrs''} \\
\end{array}
\end{array}
\]
Now, $\tbf$ iterates $\tstep$ on a non-empty forest, which is
represented by a single rose tree $r$ and a possibly empty list of
rose trees $\vrs$.  It is defined by
induction on $\vari$, the strict upper bound of the tree heights.
\[
\begin{array}{lll}
\tbf & :  & \forall \vari\forall A.\, \Rose^\vari\,A \to
\List^\infty(\Rose^\vari\,A) \to \List^\infty A \\
\tbf & := & \tfixmu_0 \lambda \vbf \lambda r \lambda \vrs.\
  \tmatchraw\ \tstep\,(\tcons\;r\,\vrs)\ \twith \\
&& \quad 
\begin{array}{lll}
  \ttuple{\vas,\ \tnil} & \tmarr & \vas \\
  \ttuple{\vas,\ \tcons\;r'\,\vrs'} & \tmarr &
    \tappend\,\vas\,(\vbf\,r'\,\vrs') \\
\end{array}
\end{array}
\]
Function $\tbf$ terminates because the recursive-call trees in forest
$\tcons\,r'\,\vrs'$ are smaller than the input trees in forest
$\tcons\,r\,\vrs$.  This information is available to the type system
through the type of $\tstep$.  The type of $\tbf$ is admissible for
recursion since $\List^\infty\,(\Rose^\vari\,A)$ is lower
semi-continuous in $\vari$---thanks to Thm.~\ref{thm:mucont} and rule $\rcontmu$.
\end{exa}
It is clear that admissibility is no way complete.  One can find
trivial examples of terminating programs which are refused by the type
system because they fail the admissibility check.  For instance, take
the recursive identity function of type $\forall \vari.\, \Nat^\vari
\to \Nat^\vari$ and add an unused argument of type $\Nat^\infty \to
\Nat^\vari$:
\[
\begin{array}{l@{~}lll}
  \multicolumn 4 l { 
    \tloopnot  : \forall \vari.\, \Nat^\vari
      \to (\Nat^\infty \to \Nat^\vari) \to \Nat^\vari 
    } \\
  \tloopnot\ 0    & g & = & 0 \\
  \tloopnot\ (n+1)& g & = & 1 + \tloopnot\,n\,(\tshift\,g) \\
\end{array}
\]
Its type is not upper
semi-continuous, but of course $\tloopnot$ is terminating.

\section{Conclusions}
\label{sec:concl}

We have motivated the importance of semi-continuity for the soundness
of type-based termination checking, explored the realm of
semi-continuous functions from ordinals to semantic types, and
developed a calculus for semi-continuous types.  
We have seen a few interesting examples involving
semi-continuous types, many more can be found in the author's thesis
\cite[Ch.~6]{abel:PhD}.   These examples cannot be handled by
type-based termination \`a la Barthe \etal\
\cite{gimenez:typeBased,bartheGregoirePastawski:tlca05,bartheGregoirePastawski:lpar06},
but our developments could be directly incorporated into their calculus.

In previous work \cite{abel:tlca03}, 
I have already presented a calculus for admissible
recursion types.  But the language had neither polymorphism,
higher-kinded types, nor semi-continuous types inside each other
($\Stream^\vari\,\Nat^\vari$).  Hughes, Pareto, and Sabry
\cite{pareto:sizedtypes} have also given criteria for admissible types
similar to ours, but more ad-hoc ones, not based on the mathematical
concept of semi-continuity.  Also, a crucial difference is that we also
treat \emph{infinitely} branching data structures.  To be fair, I
should say that their work has been a major source of inspiration for me.

As a further direction of research, I propose to develop a kinding
system where semi-continuity is first class, \ie, one can abstract
over semi-continuous constructors, and kind arrows can carry the
corresponding polarities $\plower$ or $\pupper$.  First attempts
suggest that such a calculus is not straightforward, and a more
fine-grained polarity system will be necessary.
Important is also the study of semi-continuity properties of
dependent types, in order to apply these results to type-based
termination in type-theoretic proof assistants.

\para{Acknowledgments}
I would like to thank my PhD supervisor, Martin Hofmann, for discussions
on $\Fhat$.  Thanks to John Hughes for lending his ear in difficult phases of
this work; for instance, when I was trying to prove upper
semi-continuity of inductive types but then found 
\shortversion{a}\longversion{the $\Hungry$}
counterexample.  Thanks to the anonymous referees of previous
versions of this paper who gave insightful and helpful comments.

\longversion{
\appendix

\section{Complete Specification of \Fhat}
\label{sec:appendix}

The following figures display all constructs and rules of $\Fhat$.

\begin{deffigure}{\textwidth}{\Fhat: Kinds and
constructors.\label{fig:constr}}
Syntactic categories.
\[
\def\arraystretch{1.0}
\begin{array}{lrl@{\hspace{4ex}}l}
  p & ::= & \pco \mid \pcontra \mid \pin 
    & \mbox{polarity} \\
  \kappa & ::= & \ttype \mid \tord \mid p\kappa \to \kappa' 
    & \mbox{kind} \\
  \pkind & ::= & \ttype \mid p\pkind \to \pkind'
    & \mbox{pure kind} \\
  a,b,A,B,F,G & ::= & C \mid X \mid \lambda X\ofkinddot\kappa F \mid F\,G
    & \mbox{(type) constructor} \\
  C & ::= & 1 \mid + \mid \times \mid {\to} \mid \forall_\kappa \mid
    \muka \mid \nuka \mid \s \mid \infty
    & \mbox{constructor constant} \\
  \Delta & ::= & \cempty \mid \Delta, X\of p\kappa
    & \mbox{polarized context}
\end{array}
\]
The signature $\Sigma$ assigns kinds to constants  ($\kappa \tox p
\kappa'$ means $p\kappa \to \kappa'$).
\[
\def\arraystretch{1.3}
\begin{array}[t]{lll@{\hspace{4ex}}l}
 1      & : & \ttype
    & \mbox{ unit type } \\
 +      & : & \ttype \toco \ttype \toco \ttype
    & \mbox{ disjoint sum }  \\
 \times & : & \ttype \toco \ttype \toco \ttype
    & \mbox{ cartesian product }  \\
 \mbox{$\to$} & : & \tominus \ttype {\toplus \ttype \ttype}
    & \mbox{ function space } \\
 \forall_\kappa & : & \toplus{(\tozero \kappa \ttype)}\ttype
    & \mbox{ quantification }\\
 \muka & : & \toplus \tord \kfix 
    & \mbox{ inductive constructors} \\
 \nuka & : & \tominus \tord \kfix
    & \mbox{ coinductive constructors} \\
 \s & : & \toplus \tord \tord 
    & \mbox{ successor of ordinal } \\
 \infty & : & \tord
    & \mbox{ infinity ordinal } \\
\end{array}
\]
Notation.
\begin{gather*}
  \nabla \abbrfor \mu \mbox{ or } \nu 
\qquad
  \nabla^a \abbrfor \nabla a 
\\[1ex]
  \forall X\ofkinddot\kappa A \abbrfor 
    \forall_\kappa(\lambda X\ofkinddot\kappa A)
\qquad
  \forall X A \abbrfor \forall X\ofkinddot\kappa A
\qquad
  \lambda X F \abbrfor \lambda X\ofkinddot\kappa F
\\[1ex]
  A + B \abbrfor {+}\,A\,B
\qquad
  A \times B \abbrfor {\times}\,A\,B
\qquad
  A \to B \abbrfor {\to}\,A\,B
\end{gather*}
Ordering and composition of polarities.
\[
  p \leq p \qquad
  \pin \leq p \qquad
  \pco p = p \qquad
  \pcontra \pcontra = \pco \qquad
  \pin p = \pin \qquad
  p p' = p' p
\]
Inverse application of a polarity to a context.
\[
\begin{array}[t]{lll}
  \pinv p \cempty & = & \cempty \\
  \pinv\pco\Delta & = & \Delta \\
  \pinv\pcontra(\Delta, X\ofkind{p\kappa}) & = &
  (\pinv\pcontra\Delta), X\ofkind{(\pcontra p)\kappa} 
\\
\end{array}
\qquad
\begin{array}[t]{lll}
  \pinv\pin(\Delta, X\ofkind{\pin    \kappa}) & = & (\pinv\pin\Delta), X\ofkind{\pin\kappa} \\
  \pinv\pin(\Delta, X\ofkind{\pco    \kappa}) & = & \pinv\pin\Delta \\
  \pinv\pin(\Delta, X\ofkind{\pcontra\kappa}) & = & \pinv\pin\Delta \\
\end{array}
\]
Kinding $\Delta \der F : \kappa$.
\begin{gather*}
  \nru{\rkindc
    }{C\ofkind{\kappa} \in \Sigma
    }{\Delta \der C : \kappa}
\qquad
  \nru{\rkindvar
    }{X\ofkind{p\kappa} \in \Delta \qquad p \leq \pco
    }{\Delta \der X : \kappa}
\\[2ex]
  \nru{\rkindabs
    }{\Delta, X \ofkind{p\kappa} \der F : \kappa'
    }{\Delta \der \lambda X\ofkinddot\kappa  F : p\kappa \to \kappa'}
\qquad
  \nru{\rkindapp
    }{\Delta \der F : p\kappa \to \kappa' \qquad
      \pinv p\Delta \der G : \kappa
    }{\Delta \der F\,G : \kappa'  }
\end{gather*}
\end{deffigure}

\begin{deffigure}{\textwidth}{\Fhat: Constructor equality and subtyping.\label{fig:subty}}
Constructor equality $\Delta \der F = F' : \kappa$.
\begin{gather*}
  \nru{\reqinf}{}{\Delta \derD \s\,\infty \eq \infty : \tord}
\\[2ex]
  \nru{\reqbeta
     }{\Delta, X\ofkind{p\kappa} \derD F : \kappa' \qquad
      \pinv p\Delta \derD G : \kappa
     }{\Delta \derD (\lambda X\ofkinddot\kappa  F)\, G \eq \subst G X F : \kappa'}
\qquad
  \nru{\reqeta
     }{\Delta \derD F : p\kappa\to\kappa'
     }{\Delta \derD (\lambda X\ofkinddot\kappa \,F\,X) \eq F : p\kappa\to\kappa'}
\\[2ex]
  \nru{\reqvar
     }{X\ofkind{p\kappa} \in \Delta \qquad 
       p \leq \pco
     }{\Delta \derD X \eq X : \kappa} 
\qquad
  \nru{\reqlam
    }{\Delta, X\ofkind{p\kappa} \derD F \eq  F' : \kappa'
    }{\Delta \derD \lambda X\ofkinddot\kappa  F \eq 
                   \lambda X\ofkinddot\kappa  F' : p\kappa \to \kappa' }
\\[2ex]
  \nru{\reqc
     }{C\ofkind{\kappa} \in \Sigma
     }{\Delta \derD C \eq C : \kappa}
\qquad
  \nru{\reqapp
    }{\Delta \derD F \eq F' : p\kappa \to \kappa' \qquad 
     \pinv p\Delta \derD G \eq G' : \kappa
    }{\Delta \derD F\,G \eq F'\,G' : \kappa'}
\\[2ex]
  \nru{\reqsym 
     }{\Delta \derD F \eq F' : \kappa
     }{\Delta \derD F' \eq F : \kappa} 
\qquad
  \nru{\reqtrans
     }{\Delta \derD F_1 \eq F_2 : \kappa \qquad 
      \Delta \derD F_2 \eq F_3 : \kappa
     }{\Delta \derD F_1 \eq F_3 : \kappa}
\end{gather*}
Constructor subtyping $\Delta \der F \leq F' : \kappa$.  
\begin{gather*}
  \nru{\rdleqsr
     }{\Delta \der a : \tord
     }{\Delta \der a \leq \s\,a : \tord}
\qquad
  \nru{\rdleqinf}{\Delta \der a : \tord}{\Delta \der a \leq \infty :
    \tord}
\\[2ex]
  \nru{\rdleqlam
     }{\Delta,\, X\ofkind{p\kappa} \der F \leq F' : \kappa'
     }{\Delta \derD \lambda X\ofkinddot\kappa  F 
               \leq \lambda X\ofkinddot\kappa  F' : p\kappa \to \kappa'}
\\[2ex]
  \nru{\rdleqapp
     }{\Delta \derD F \leq F' : p\kappa \to \kappa' \qquad
       \pinv p\Delta \derD G : \kappa
     }{\Delta \derD F\,G \leq F'\,G : \kappa'}
\\[2ex]
  \nru{\rdleqappco
     }{\Delta \derD F : \pco\kappa \to \kappa' \qquad
       \Delta \derD G \leq G' : \kappa
     }{\Delta \derD F\,G \leq F\,G' : \kappa'}
\\[2ex]
  \nru{\rdleqappcontra
     }{\Delta \derD F : \pcontra\kappa \to \kappa' \qquad
       \pinv\pcontra\Delta \derD G' \leq G : \kappa
     }{\Delta \derD F\,G \leq F\,G' : \kappa'}  
\\[2ex]
  \nru{\rdleqrefl
     }{\Delta \derD F \eq F' : \kappa
     }{\Delta \derD F \leq F' : \kappa}
\qquad
  \nru{\rdleqtrans
     }{\Delta \derD F_1 \leq F_2 : \kappa \qquad
       \Delta \derD F_2 \leq F_3 : \kappa
     }{\Delta \derD F_1 \leq F_3 : \kappa}
\\[2ex]
  \nru{\rdleqasymm
     }{\Delta \derD F \leq F' : \kappa \qquad
       \Delta \derD F' \leq F : \kappa
     }{\Delta \derD F \eq F' : \kappa}
\end{gather*}
\end{deffigure}

\begin{deffigure}{\textwidth}{\Fhat: Terms, reduction and typing.\label{fig:termty}}
Syntactic categories.  
\[
\begin{array}{lrl@{\hspace{4ex}}l}
  r, s, t & ::= & c \mid x \mid \lambda x t \mid r\,s 
  & \mbox{term} 
\\
  c & ::= & \ttuple{} \mid \tpair \mid \tfst \mid \tsnd
    \mid \tinl \mid \tinr \mid \tcaseraw \mid \tin \mid \tout
    \mid \tfixmu_n \mid \tfixnu_n 
  & \mbox{constant ($n\in\NN$)}
\\
  v & ::= & \lambda x t \mid \ttuple{} \mid \tpair\,t_1\,t_2 \mid \tinl\,t
                     \mid \tinr\,t \mid \tin\,t \mid c \mid \tpair\,t 
             \mid \tfixnabla_n s\,t_{1..m}
    & \mbox{value ($m<n$)}
\\
  e(\_) & ::= & \_\,s \mid \tfst\,\_ \mid \tsnd\,\_ \mid \tcaseraw\,\_
                \mid \tout\,\_ \mid \tfixmu_n\, s\ t_{1..n}\, \_
    & \mbox{evaluation frame} \\
  E(\_) & ::= & e_1(\dots e_n(\_)\dots)
    & \mbox{eval.\ cxt.$ (n \geq 0)$}\\
  \Gamma & ::= & \cempty \mid \Gamma, x\of A \mid \Gamma,
    X\ofkind{p\kappa}
  & \mbox{typing context}
\\
\end{array}
\]
Notation:
\[
  \ttuple{r,s} \abbrfor \tpair\,r\,s \qquad
  t_{1..n} \abbrfor t_1\,t_2\dots t_n.
\]
Reduction $t \red t'$.
\[
\begin{array}[t]{lll@{\hspace{4ex}}l}
  (\lambda x t)\, s & \red & \subst s x t \\
  \tfst\,\ttuple{r,s} & \red & r \\
  \tsnd\,\ttuple{r,s} & \red & s \\
  \tcaseraw\,(\tinl\,r) & \red & \lambda x\lambda y.\,x\,r \\
  \tcaseraw\,(\tinr\,r) & \red & \lambda x\lambda y.\,y\,r \\
\end{array}
\qquad
\begin{array}[t]{lll@{\hspace{4ex}}l}
  \tout\,(\tin\,r) & \red & r \\
  \tfixmu_n\, s\, t_{1..n}\, (\tin\,t) & \red & 
    s\, (\tfixmu_n\, s)\, t_{1..n}\, (\tin\,t) \\
  \tout\,(\tfixnu_n\, s\, t_{1..n}) & \red & 
    \tout\,(s\, (\tfixnu_n\, s)\, t_{1..n})
  \\
  \\
  \subst s x t & \red & \subst {s'}x t \quad \mbox{ if } s \red s' \\
\end{array}
\]
The signature $\Sigma$ contains types for some constants:
\begin{gather*}
\begin{array}{lll@{\hspace{4ex}}l}
  \tpair & : & \forall A \forall B.\ A \to B \to A \times B \\
  \tfst  & : & \forall A \forall B.\ A \times B \to A \\
  \tsnd  & : & \forall A \forall B.\ A \times B \to B \\
\end{array}
\qquad
\begin{array}{lll@{\hspace{4ex}}l}
  \ttuple{} & : & 1 \\
  \tinl  & : & \forall A \forall B.\ A \to A + B \\
  \tinr  & : & \forall A \forall B.\ B \to A + B \\
\end{array}
\\[1ex]
\begin{array}{lll@{\hspace{4ex}}l}
  \tcaseraw & : & \forall A \forall B \forall C.\ A + B \to (A \to C) \to
  (B \to C) \to C \\
  \tin   & : & \forall F\ofkinddot{\kappa\toco\kappa}
               \forall G^1\ofkind{\pkind^1} \dots
               \forall G^n\ofkinddot{\pkind^n}
               \forall \vari\ofkinddot\tord
               F\,(\nablakappa^\vari F)\,\vec G \to 
               \nablakappa^{\vari+1}F\,\vec G
\\
  \tout  & : & \forall F\ofkinddot{\kappa\toco\kappa}
               \forall G^1\ofkind{\pkind^1} \dots
               \forall G^n\ofkinddot{\pkind^n}
               \forall \vari\ofkinddot\tord
               \nablakappa^{\vari+1}F\,\vec G \to 
               F\,(\nablakappa^\vari F)\,\vec G 
\\
   && \qquad (\nabla \in \{ \mu, \nu \},\ \kappa = \vec\pkind\tox{\vec p}\ttype)
\end{array}
\end{gather*}
Well-formed typing contexts.
\begin{gather*}
  \nru{\rcxtempty}{}{\cempty \cxt}
\quad
  \nru{\rcxttyvar
     }{\Gamma \cxt
     }{\Gamma, X\ofkind{\pin\kappa} \cxt}
\quad
  \nru{\rcxtvar
     }{\Gamma \cxt \qquad \Gamma \der A : \ttype
     }{\Gamma, x\of A \cxt}
\end{gather*}
Typing $\Gamma \der t : A$.
\begin{gather*}
  \nru{\rtyc
     }{(c \of A) \in \Sigma
     }{\Gamma \der c : A}
\qquad
  \nru{\rtyvar
     }{(x\of A) \in \Gamma \qquad \Gamma \cxt
     }{\Gamma \der x : A}
\qquad
  \nru{\rtyabs
     }{\Gamma, x \of A \der t : B
     }{\Gamma \der \lambda x t : A \to B}
\\[2ex]
  \nru{\rtyapp
     }{\Gamma \der r : A \to B \qquad
       \Gamma \der s : A
     }{\Gamma \der r\,s : B}
\qquad
  \nru{\rtysub}{\Gamma \der t : A \qquad
       \Gamma \der A \leq B : \ttype
     }{\Gamma \der t : B}
\\[2ex]
  \nru{\rtygen}{\Gamma, X\ofkind{\pin\kappa} \der t : F\,X
     }{\Gamma \der t : \forallk{\kappa} F}
\qquad
  \nru{\rtyinst}{\Gamma \der t\ :\ \forallk{\kappa}\, F \qquad
       \Gamma \der G : \kappa
     }{\Gamma \der t : F\,G}
\\[2ex] 
\nru{\rtyrec
   }{\Gamma \der A \adm{\tfixnabla_n\!} \qquad
     \Gamma \der a : \tord 
   }{\Gamma \der \tfixnabla_n : (\forall \vari\of\tord.\,
     A\,\vari\to A\,(\vari+1)) \to A\,a}
\end{gather*}
\end{deffigure}

}%

\clearpage

\bibliographystyle{alpha}
\bibliography{all}

\end{document}

%% file: prooftree.tex
\newdimen\proofrulebreadth \proofrulebreadth=.05em
\newdimen\proofdotseparation \proofdotseparation=1.25ex
\newdimen\proofrulebaseline \proofrulebaseline=2ex
\newcount\proofdotnumber \proofdotnumber=3
\let\then\relax
\def\hfi{\hskip0pt plus.0001fil}
\mathchardef\squigto="3A3B
\newif\ifinsideprooftree\insideprooftreefalse
\newif\ifonleftofproofrule\onleftofproofrulefalse
\newif\ifproofdots\proofdotsfalse
\newif\ifdoubleproof\doubleprooffalse
\let\wereinproofbit\relax
\newdimen\shortenproofleft
\newdimen\shortenproofright
\newdimen\proofbelowshift
\newbox\proofabove
\newbox\proofbelow
\newbox\proofrulename
\def\shiftproofbelow{\let\next\relax\afterassignment\setshiftproofbelow\dimen0 }
\def\shiftproofbelowneg{\def\next{\multiply\dimen0 by-1 }%
\afterassignment\setshiftproofbelow\dimen0 }
\def\setshiftproofbelow{\next\proofbelowshift=\dimen0 }
\def\setproofrulebreadth{\proofrulebreadth}

\def\prooftree{
\ifnum  \lastpenalty=1
\then   \unpenalty
\else   \onleftofproofrulefalse
\fi
\ifonleftofproofrule
\else   \ifinsideprooftree
        \then   \hskip.5em plus1fil
        \fi
\fi
\bgroup
\setbox\proofbelow=\hbox{}\setbox\proofrulename=\hbox{}%
\let\justifies\proofover\let\leadsto\proofoverdots\let\Justifies\proofoverdbl
\let\using\proofusing\let\[\prooftree
\ifinsideprooftree\let\]\endprooftree\fi
\proofdotsfalse\doubleprooffalse
\let\thickness\setproofrulebreadth
\let\shiftright\shiftproofbelow \let\shift\shiftproofbelow
\let\shiftleft\shiftproofbelowneg
\let\ifwasinsideprooftree\ifinsideprooftree
\insideprooftreetrue
\setbox\proofabove=\hbox\bgroup$\displaystyle 
\let\wereinproofbit\prooftree
\shortenproofleft=0pt \shortenproofright=0pt \proofbelowshift=0pt
\onleftofproofruletrue\penalty1
}

\def\eproofbit{
\ifx    \wereinproofbit\prooftree
\then   \ifcase \lastpenalty
        \then   \shortenproofright=0pt  
        \or     \unpenalty\hfil         
        \or     \unpenalty\unskip       
        \else   \shortenproofright=0pt  
        \fi
\fi
\global\dimen0=\shortenproofleft
\global\dimen1=\shortenproofright
\global\dimen2=\proofrulebreadth
\global\dimen3=\proofbelowshift
\global\dimen4=\proofdotseparation
\global\count255=\proofdotnumber
$\egroup  
\shortenproofleft=\dimen0
\shortenproofright=\dimen1
\proofrulebreadth=\dimen2
\proofbelowshift=\dimen3
\proofdotseparation=\dimen4
\proofdotnumber=\count255
}

\def\proofover{
\eproofbit 
\setbox\proofbelow=\hbox\bgroup 
\let\wereinproofbit\proofover
$\displaystyle
}%
\def\proofoverdbl{
\eproofbit 
\doubleprooftrue
\setbox\proofbelow=\hbox\bgroup 
\let\wereinproofbit\proofoverdbl
$\displaystyle
}%
\def\proofoverdots{
\eproofbit 
\proofdotstrue
\setbox\proofbelow=\hbox\bgroup 
\let\wereinproofbit\proofoverdots
$\displaystyle
}%
\def\proofusing{
\eproofbit 
\setbox\proofrulename=\hbox\bgroup 
\let\wereinproofbit\proofusing
\kern0.3em$
}

\def\endprooftree{
\eproofbit 
  \dimen5 =0pt
\dimen0=\wd\proofabove \advance\dimen0-\shortenproofleft
\advance\dimen0-\shortenproofright
\dimen1=.5\dimen0 \advance\dimen1-.5\wd\proofbelow
\dimen4=\dimen1
\advance\dimen1\proofbelowshift \advance\dimen4-\proofbelowshift
\ifdim  \dimen1<0pt
\then   \advance\shortenproofleft\dimen1
        \advance\dimen0-\dimen1
        \dimen1=0pt
        \ifdim  \shortenproofleft<0pt
        \then   \setbox\proofabove=\hbox{%
                        \kern-\shortenproofleft\unhbox\proofabove}%
                \shortenproofleft=0pt
        \fi
\fi
\ifdim  \dimen4<0pt
\then   \advance\shortenproofright\dimen4
        \advance\dimen0-\dimen4
        \dimen4=0pt
\fi
\ifdim  \shortenproofright<\wd\proofrulename
\then   \shortenproofright=\wd\proofrulename
\fi
\dimen2=\shortenproofleft \advance\dimen2 by\dimen1
\dimen3=\shortenproofright\advance\dimen3 by\dimen4
\ifproofdots
\then
        \dimen6=\shortenproofleft \advance\dimen6 .5\dimen0
        \setbox1=\vbox to\proofdotseparation{\vss\hbox{$\cdot$}\vss}%
        \setbox0=\hbox{%
                \advance\dimen6-.5\wd1
                \kern\dimen6
                $\vcenter to\proofdotnumber\proofdotseparation
                        {\leaders\box1\vfill}$%
                \unhbox\proofrulename}%
\else   \dimen6=\fontdimen22\the\textfont2 
        \dimen7=\dimen6
        \advance\dimen6by.5\proofrulebreadth
        \advance\dimen7by-.5\proofrulebreadth
        \setbox0=\hbox{%
                \kern\shortenproofleft
                \ifdoubleproof
                \then   \hbox to\dimen0{%
                        $\mathsurround0pt\mathord=\mkern-6mu%
                        \cleaders\hbox{$\mkern-2mu=\mkern-2mu$}\hfill
                        \mkern-6mu\mathord=$}%
                \else   \vrule height\dimen6 depth-\dimen7 width\dimen0
                \fi
                \unhbox\proofrulename}%
        \ht0=\dimen6 \dp0=-\dimen7
\fi
\let\doll\relax
\ifwasinsideprooftree
\then   \let\VBOX\vbox
\else   \ifmmode\else$\let\doll=$\fi
        \let\VBOX\vcenter
\fi
\VBOX   {\baselineskip\proofrulebaseline \lineskip.2ex
        \expandafter\lineskiplimit\ifproofdots0ex\else-0.6ex\fi
        \hbox   spread\dimen5   {\hfi\unhbox\proofabove\hfi}%
        \hbox{\box0}%
        \hbox   {\kern\dimen2 \box\proofbelow}}\doll%
\global\dimen2=\dimen2
\global\dimen3=\dimen3
\egroup 
\ifonleftofproofrule
\then   \shortenproofleft=\dimen2
\fi
\shortenproofright=\dimen3
\onleftofproofrulefalse
\ifinsideprooftree
\then   \hskip.5em plus 1fil \penalty2
\fi
}


%% file: lmcs08.bbl
\newcommand{\etalchar}[1]{$^{#1}$}
\begin{thebibliography}{BFG{\etalchar{+}}04}

\bibitem[Abe03]{abel:tlca03}
Andreas Abel.
\newblock Termination and guardedness checking with continuous types.
\newblock In M.~Hofmann, editor, {\em Typed Lambda Calculi and Applications
  (TLCA 2003), Valencia, Spain}, volume 2701 of {\em Lecture Notes in Computer
  Science}, pages 1--15. Springer-Verlag, June 2003.

\bibitem[Abe04]{abel:rairo04}
Andreas Abel.
\newblock Termination checking with types.
\newblock {\em RAIRO -- Theoretical Informatics and Applications},
  38(4):277--319, 2004.
\newblock Special Issue: Fixed Points in Computer Science (FICS'03).

\bibitem[Abe06a]{abel:csr06}
Andreas Abel.
\newblock Polarized subtyping for sized types.
\newblock In Dima Grigoriev, John Harrison, and Edward~A. Hirsch, editors, {\em
  Computer Science - Theory and Applications, First International Computer
  Science Symposium in Russia, CSR 2006, St. Petersburg, Russia, June 8-12,
  2006, Proceedings}, volume 3967 of {\em Lecture Notes in Computer Science},
  pages 381--392. Springer-Verlag, 2006.

\bibitem[Abe06b]{abel:PhD}
Andreas Abel.
\newblock {\em A Polymorphic Lambda-Calculus with Sized Higher-Order Types}.
\newblock PhD thesis, Ludwig-Maximilians-Universit\"at M\"unchen, 2006.

\bibitem[Abe06c]{abel:csl06}
Andreas Abel.
\newblock Semi-continuous sized types and termination.
\newblock In Zolt\'an \'Esik, editor, {\em Computer Science Logic, 20th
  International Workshop, CSL 2006, 15th Annual Conference of the EACSL,
  Szeged, Hungary, September 21-24, 2006, Proceedings}, volume 4207 of {\em
  Lecture Notes in Computer Science}, pages 72--88. Springer-Verlag, 2006.

\bibitem[ACG98]{amadio:guardcondition}
Roberto~M. Amadio and Solange Coupet-Grimal.
\newblock Analysis of a guard condition in type theory (extended abstract).
\newblock In Maurice Nivat, editor, {\em Foundations of Software Science and
  Computation Structure, First International Conference, FoSSaCS'98, Held as
  Part of the European Joint Conferences on the Theory and Practice of
  Software, ETAPS'98, Lisbon, Portugal, March 28 - April 4, 1998, Proceedings},
  volume 1378 of {\em Lecture Notes in Computer Science}, pages 48--62.
  Springer-Verlag, 1998.

\bibitem[Alt01]{alti:coinductiveRep}
Thorsten Altenkirch.
\newblock Representations of first order function types as terminal coalgebras.
\newblock In Samson Abramsky, editor, {\em Fifth International Conference on
  Typed Lambda Calculi and Applications}, volume 2044 of {\em Lecture Notes in
  Computer Science}, pages 8--21. Springer-Verlag, 2001.

\bibitem[AM04]{abelMatthes:csl04}
Andreas Abel and Ralph Matthes.
\newblock Fixed points of type constructors and primitive recursion.
\newblock In Jerzy Marcinkowski and Andrzej Tarlecki, editors, {\em Computer
  Science Logic, 18th International Workshop, CSL 2004, 13th Annual Conference
  of the EACSL, Karpacz, Poland, September 20-24, 2004, Proceedings}, volume
  3210 of {\em Lecture Notes in Computer Science}, pages 190--204.
  Springer-Verlag, 2004.

\bibitem[AR99]{alti:monadic}
Thorsten Altenkirch and Bernhard Reus.
\newblock Monadic presentations of lambda terms using generalized inductive
  types.
\newblock In J{\"o}rg Flum and Mario Rodr{\'\i}guez-Artalejo, editors, {\em
  Computer Science Logic, 13th International Workshop, CSL '99, 8th Annual
  Conference of the EACSL, Madrid, Spain, September 20-25, 1999, Proceedings},
  volume 1683 of {\em Lecture Notes in Computer Science}, pages 453--468.
  Springer-Verlag, 1999.

\bibitem[BFG{\etalchar{+}}04]{gimenez:typeBased}
Gilles Barthe, Maria~J. Frade, Eduardo Gim\'enez, Luis Pinto, and Tarmo
  Uustalu.
\newblock Type-based termination of recursive definitions.
\newblock {\em Mathematical Structures in Computer Science}, 14(1):97--141,
  2004.

\bibitem[BGP05]{bartheGregoirePastawski:tlca05}
Gilles Barthe, Benjamin Gr{\'e}goire, and Fernando Pastawski.
\newblock Practical inference for type-based termination in a polymorphic
  setting.
\newblock In Pawel Urzyczyn, editor, {\em Typed Lambda Calculi and Applications
  (TLCA 2005), Nara, Japan}, volume 3461 of {\em Lecture Notes in Computer
  Science}, pages 71--85. Springer-Verlag, 2005.

\bibitem[BGP06]{bartheGregoirePastawski:lpar06}
Gilles Barthe, Benjamin Gr{\'e}goire, and Fernando Pastawski.
\newblock {CIC\^{}}: Type-based termination of recursive definitions in the
  {C}alculus of {I}nductive {C}onstructions.
\newblock In Miki Hermann and Andrei Voronkov, editors, {\em Logic for
  Programming, Artificial Intelligence, and Reasoning, 13th International
  Conference, LPAR 2006, Phnom Penh, Cambodia, November 13-17, 2006,
  Proceedings}, volume 4246 of {\em Lecture Notes in Computer Science}, pages
  257--271. Springer-Verlag, 2006.

\bibitem[Bla04]{blanqui:rta04}
Fr\'ed\'eric Blanqui.
\newblock A type-based termination criterion for dependently-typed higher-order
  rewrite systems.
\newblock In Vincent van Oostrom, editor, {\em Rewriting Techniques and
  Applications, 15th International Conference, RTA 2004, Aachen, Germany, June
  3 -- 5, 2004, Proceedings}, volume 3091 of {\em Lecture Notes in Computer
  Science}, pages 24--39. Springer-Verlag, 2004.

\bibitem[Bla05]{blanqui:csl05}
Fr{\'e}d{\'e}ric Blanqui.
\newblock Decidability of type-checking in the {Calculus of Algebraic
  Constructions} with size annotations.
\newblock In C.-H.~Luke Ong, editor, {\em Computer Science Logic, 19th
  International Workshop, CSL 2005, 14th Annual Conference of the EACSL,
  Oxford, UK, August 22-25, 2005, Proceedings}, volume 3634 of {\em Lecture
  Notes in Computer Science}, pages 135--150. Springer-Verlag, 2005.

\bibitem[BP99]{bird:debruijn}
Richard~S. Bird and Ross Paterson.
\newblock De {B}ruijn notation as a nested datatype.
\newblock {\em Journal of Functional Programming}, 9(1):77--91, 1999.

\bibitem[CW99]{crary:lx}
Karl Crary and Stephanie Weirich.
\newblock Flexible type analysis.
\newblock In {\em Proceedings of the fourth ACM SIGPLAN International
  Conference on Functional Programming (ICFP '99), Paris, France}, volume~34 of
  {\em SIGPLAN Notices}, pages 233--248. ACM Press, 1999.

\bibitem[DC99]{dugganCompagnoni:subtyping}
Dominic Duggan and Adriana Compagnoni.
\newblock Subtyping for object type constructors, January 1999.
\newblock Presented at FOOL 6.

\bibitem[Gim98]{gimenez:strec}
Eduardo Gim\'enez.
\newblock Structural recursive definitions in type theory.
\newblock In K.~G. Larsen, S.~Skyum, and G.~Winskel, editors, {\em Automata,
  Languages and Programming, 25th International Colloquium, ICALP'98, Aalborg,
  Denmark, July 13-17, 1998, Proceedings}, volume 1443 of {\em Lecture Notes in
  Computer Science}, pages 397--408. Springer-Verlag, 1998.

\bibitem[Han02]{hancock:numberclasses}
Peter Hancock.
\newblock The step to the next number class.
\newblock http://www.dcs.ed.ac.uk/home/pgh/number-classes.html, 2002.

\bibitem[Hin00a]{hinze:efficientGfolds}
Ralf Hinze.
\newblock Efficient generalized folds.
\newblock In Johan Jeuring, editor, {\em Proceedings of the Second Workshop on
  Generic Programming, WGP 2000}, Ponte de Lima, Portugal, July 2000.

\bibitem[Hin00b]{hinze:GGTries}
Ralf Hinze.
\newblock Generalizing generalized tries.
\newblock {\em Journal of Functional Programming}, 10(4):327--351, July 2000.

\bibitem[HPS96]{pareto:sizedtypes}
John Hughes, Lars Pareto, and Amr Sabry.
\newblock Proving the correctness of reactive systems using sized types.
\newblock In {\em 23rd ACM SIGPLAN-SIGACT Symposium on Principles of
  Programming Languages, POPL'96}, pages 410--423, 1996.

\bibitem[Men87]{mendler:lics}
Nax~Paul Mendler.
\newblock Recursive types and type constraints in second-order lambda calculus.
\newblock In {\em Proceedings of the Second Annual IEEE Symposium on Logic in
  Computer Science, Ithaca, N.Y.}, pages 30--36. IEEE Computer Society Press,
  1987.

\bibitem[Par00]{pareto:PhD}
Lars Pareto.
\newblock {\em Types for Crash Prevention}.
\newblock PhD thesis, Chalmers University of Technology, 2000.

\bibitem[Pie02]{pierce:tapl}
Benjamin~C. Pierce.
\newblock {\em Types and Programming Languages}.
\newblock MIT Press, 2002.

\bibitem[PM92]{paulin:inductiveTR}
Christine Paulin-Mohring.
\newblock Inductive definitions in the system {Coq}---rules and properties.
\newblock Technical report, Laboratoire de l'Informatique du Parall\'elisme,
  December 1992.

\bibitem[Ste98]{steffen:PhD}
Martin Steffen.
\newblock {\em Polarized Higher-Order Subtyping}.
\newblock PhD thesis, Technische Fakult{\"a}t, Universit{\"a}t Erlangen, 1998.

\bibitem[Vou04]{vouillon:subtypingUnion}
J\'er\^ome Vouillon.
\newblock Subtyping union types.
\newblock In Jerzy Marcinkowski and Andrzej Tarlecki, editors, {\em Computer
  Science Logic, 18th International Workshop, CSL 2004, 13th Annual Conference
  of the EACSL, Karpacz, Poland, September 20-24, 2004, Proceedings}, volume
  3210 of {\em Lecture Notes in Computer Science}, pages 415--429.
  Springer-Verlag, 2004.

\bibitem[Xi01]{xi:termination}
Hongwei Xi.
\newblock Dependent types for program termination verification.
\newblock In {\em Proceedings of 16th IEEE Symposium on Logic in Computer
  Science}, Boston, USA, June 2001.

\end{thebibliography}
